\documentclass[fleqn,usenatbib]{mnras}

\usepackage{newtxtext,newtxmath}

\usepackage[T1]{fontenc}

\DeclareRobustCommand{\VAN}[3]{#2}
\let\VANthebibliography\thebibliography
\def\thebibliography{\DeclareRobustCommand{\VAN}[3]{##3}\VANthebibliography}

\usepackage{graphicx}	
\usepackage{amsmath}	

\usepackage{orcidlink}

\title[No hidden AGN in massive PSBs]{No Hidden Monsters: Probing recently quenched galaxies for obscured AGN with \emph{JWST}-PRIMER MIRI and NIRCam}

\author[G. Hewitt et al.]{Guillaume Hewitt$^{1}$\thanks{E-mail: guillaume.hewitt@nottingham.ac.uk}\orcidlink{0009-0006-7827-007X}, Omar Almaini$^{1}$\orcidlink{0000-0001-9328-3991}, David Maltby$^{1}$\orcidlink{0000-0002-8163-080X}, Emma Chapman$^{1}$\orcidlink{0000-0002-5050-9847}, Thomas de Lisle$^{1}$\orcidlink{0009-0004-1726-0260},
\newauthor Pallavi Patil$^{2}$, Kate Rowlands$^{3,2}$\orcidlink{0000-0001-7883-8434}, Maya Skarbinski$^{2}$\orcidlink{0009-0004-0844-0657}, Elizabeth Taylor$^{4}$\orcidlink{0000-0001-8728-2984}, Vivienne Wild$^{5}$\orcidlink{0000-0002-8956-7024},
\newauthor Adam C. Carnall$^{4}$\orcidlink{0000-0002-1482-5818}, James S. Dunlop$^{4}$\orcidlink{0000-0002-1404-5950}, Norman Grogin$^{3}$, Anton M. Koekemoer$^{3}$\orcidlink{0000-0002-6610-2048}, Derek J. McLeod$^{4}$\orcidlink{0000-0003-4368-3326},
\newauthor Pablo G. Pérez-González$^{6}$\orcidlink{0000-0003-4528-5639}
\\
$^{1}$ School of Physics and Astronomy, University of Nottingham, Nottingham, NG7 2RD, U.K \\
$^{2}$ William H. Miller III Department of Physics and Astronomy, Johns Hopkins University, Baltimore, MD 21218, USA \\
$^{3}$ Space Telescope Science Institute, 3700 San Martin Drive, Baltimore, MD 21218, USA \\
$^{4}$ Institute for Astronomy, University of Edinburgh, Royal Observatory, Edinburgh, EH9 3HJ, UK \\
$^{5}$ School of Physics and Astronomy, University of St Andrews, North Haugh, St Andrews, KY16 9SS, UK \\
$^{6}$ Centro de Astrobiolog\'{\i}a (CAB/CSIC-INTA), Ctra. de Ajalvir km 4, Torrej\'on de Ardoz, E-28850, Madrid, Spain
}

\date{Accepted XXX. Received YYY; in original form ZZZ}

\pubyear{\the\year{}}

\begin{document}
\label{firstpage}
\pagerange{\pageref{firstpage}--\pageref{lastpage}}
\maketitle

\begin{abstract}
We investigate the role of obscured active galactic nuclei (AGN) in recently quenched post-starburst galaxies (PSBs), using a sample of 65 photometrically selected PSBs in the PRIMER-UDS field at $1<z<2$. Combining \emph{JWST}/MIRI 7.7 $\mu$m and 18 $\mu$m (F770W and F1800W) imaging with eight NIRCam and three \emph{HST}/ACS bands, we probe hot dust emission to test for hidden AGN or dust-enshrouded star formation. We find strong differences between the low- and high-mass PSBs. Most high-mass PSBs ($>10^{10}\textrm{\;M}_\odot$) show no excess infrared emission (consistent with the quiescent population), indicating little or no dust-obscured activity, while low-mass PSBs display enhanced emission at 18 $\mu$m, which we attribute to residual star formation. AGN template modelling indicates that the absence of mid-IR excess in massive PSBs limits any dust-enshrouded AGN to Eddington ratios of $<1\%$. In addition, we show that the F770W--F1800W colour alone is a highly effective diagnostic for separating passive and star-forming galaxies, particularly at high stellar masses. Overall, our results provide further evidence for distinct quenching pathways within the PSB population, and confirm that massive PSBs show no evidence for excess AGN activity relative to older passive galaxies.
\end{abstract}




\section{Introduction}\label{sec:Intro}
In the Universe we observe a strong bimodality in the properties of galaxies, within their colour \citep{Strateva2001,Baldry2004}, star formation rate \citep{Elbaz2007,McGee2011}, and their morphologies \citep{Driver2006,Mignoli2009}. The population within the \emph{blue cloud} generally consists of blue, star-forming galaxies with late-type (i.e. spiral) morphologies. The population within the \emph{red sequence} generally consists of `red and dead' passive galaxies, with early-type (i.e elliptical and S0) morphologies. The transformation process of galaxies from actively star-forming to quiescent (quenching) has been found to occur from well before cosmic noon through to the local Universe, with large-scale studies characterizing the buildup of the passive population through time \citep{Ilbert2010,Muzzin2013,McLeod2021}. The exact processes that drive quenching are not fully understood, and is a major focus of current galaxy research.

Seminal works exploring quenching processes over the past few decades have shown that quenching is both strongly dependent on a galaxy's mass \citep{Kauffmann2003,Bundy2006,Ilbert2013}, and its local environment \citep{Kauffmann2004,Balogh2004,Baldry2006,Muzzin2012}. Specific mechanisms have been proposed as drivers for each dependency. Processes dependent on a galaxy's mass (`mass-quenching') are attributed to internal feedback mechanisms, such as active galactic nuclei (AGN) \citep{SilkRees1998,Croton2006,Fabian2012} or stellar feedback winds and supernovae \citep{DekelSilk1986,WhiteFrenk1991,Hopkins2012}. These mechanisms can either drive direct gas removal from the galaxy, or induce heat and turbulence to halt further cooling. Processes dependent on a galaxy's environment (`environmental-quenching') are attributed to external mechanisms, either with gravitational or tidal interactions with neighbouring galaxies (i.e. mergers or harassment) \citep{Moore1996,Moore1998,Smith2015}, or through ram-pressure stripping or starvation \citep{GunnGott1972,Larson1980,Balogh2000,Peng2015}, which remove either the internal cold gas, or the hotter outer gas halo, respectively. 

The build-up of the red sequence population is due to galaxies transitioning from the blue cloud, with the lack of a major transitionary population indicating that the quenching of massive galaxies occurs on a fairly rapid timescale. To better constrain the efficacy of various quenching mechanisms, it is essential to probe the galaxies that are in the midst of transitioning. Post-starburst galaxies (PSBs) are part of this population, a rare group of galaxies believed to have been recently and rapidly quenched, following a major burst of star formation in their last Gyr. PSBs have gained increased interest over the past decade, as it is believed 25-50$\%$ of all quenching galaxies (at $z\sim1$) have gone through this phase \citep{Wild2020}. They can be identified through their characteristic spectral energy distribution (SED) profile, with distinctly strong Balmer absorption lines and the general lack of nebular emission lines, which indicates the dominance of A and F-type stars, as well as the lack of on-going star formation, respectively \citep{Dressler1983,Wild2020,French2021}. While the exact selection criteria can change depending on the particular subset of PSBs that are being probed (i.e. their exact quenching mechanism and pathway), the spectroscopic criteria for massive PSBs typically require equivalent widths of H$\delta\geq5$\AA\ and [OII] $>-5$\AA\ \citep{Balogh1999,Tran2003,Goto2007}.

While the spectroscopic criteria is effective at separating out PSBs from star-forming and quiescent galaxies, the rarity of the population makes it difficult to build up a significant spectroscopically confirmed sample. Therefore, two photometric methods have often been utilised to identify greater numbers of PSBs. The first is using a \emph{UVJ}-selection method, which builds on the widely-used colour cuts that separate star-forming and quiescent galaxies. Within the prescribed quiescent region a linear trend of stellar age has been found, leading to the ability to split the region into `young' and `old' quiescent populations \citep{Whitaker2012,Belli2019}, with the young population consisting of PSBs. Additional simulation modelling has confirmed that galaxies that pass through the PSB region have had an intense burst of star formation and rapid quenching, while those that lack a burst, and quench more gradually, move from the star-forming directly into the `old' quiescent region \citep[e.g.][]{Carnall2019,Akins2022}.

The other photometric method uses a principal component analysis (PCA) to classify galaxies based on their broad-band SEDs. The method, described in \citet{Wild2014}, finds that with just three shape parameters (which they denote as `super-colours'; SCs), star-forming, quiescent, and PSB galaxies (along with other rare populations) can be efficiently separated. These photometric classes are confirmed to be in good agreement with spectroscopic classifications \citep[][de Lisle \emph{submitted}]{Maltby2016,Wild2020}. This method has distinct advantages due to its ability to probe a wide wavelength baseline, thus being able to isolate and identify the `A-star peak' in $F_\lambda$ that is characteristic for the PSB population, as well as the strength of the 4000\,\AA/Balmer break, which can break degeneracies with dusty star-forming galaxies. The large samples of PSBs identified by this method have shown that distinct PSB populations are prevalent at different epochs. At high-redshift ($z\gtrsim2$), PSBs are massive, compact, and have a mass function structure similar to quiescent galaxies, while at low-redshift ($z\lesssim1$), PSBs have lower masses, are less compact, have a significant disc component, and have a mass function similar to the star-forming population \citep{Wild2016,Maltby2018}. This suggests that PSBs have dual quenching pathways, with the massive population dominated by internal mass-quenching mechanisms, and the low-mass population dominated by external environmental-quenching mechanisms.

Historically, the mechanism frequently invoked to drive the quenching of massive PSBs is quasar-mode AGN feedback, triggered by major gas-rich mergers \citep{SilkRees1998,Croton2006,Bower2006,Harrison2017,Belli2019,Ellison2024,Ellison2025}, which would also explain their disrupted and compact structure \citep{Almaini2017, Maltby2018}. Simulations suggest that a characteristic delay should exist between the peak of the starburst and the peak of the black hole accretion, leading to enhanced AGN activity in the PSB phase \citep{Hopkins2012AGN}. However, recent work by \citet{Almaini2025} found that massive PSBs exhibit diminished X-ray emission when compared to the star-forming population, with levels instead consistent with the passive population. The observed lack of excess AGN emission could be explained by a variety of solutions. One could be that AGN are in fact \emph{not} the primary quenching mechanism, and if present are simply `along for the ride' \citep{Lanz2022}. Another solution could be that the visibility timescale presented in \citet{Hopkins2012AGN} may not fully capture the complexity of the AGN duty cycle, with activity proceeding not as a single monolithic burst but through short, stochastic episodes consistent with the maintenance-mode models discussed in \citet{French2023} and \citet{Almaini2025}. This would allow for the retention of the outflow signatures that are observed in PSB spectra \citep{Tremonti2007,Maltby2019,Taylor2024}, which would remain visible for roughly an order of magnitude longer than the direct AGN signature.

An alternative solution could be that the AGN in massive PSBs are heavily obscured, with their X-ray and UV/optical emission being absorbed and reddened, respectively, and subsequently reprocessed at infrared wavelengths. Some models have suggested that the growth phase of SMBHs should be obscured, especially at $z>1$ \citep{Fabian1999}. Indeed, observations have shown that the majority of AGN are Type 2 and heavily obscured \citep{Risaliti1999,Lawrence2010}, with this fraction potentially increasing with redshift \citep{Hasinger2008}. Evidence that the SMBHs continue to accrete matter at a significant rate hundreds of Myrs after the starburst that activated them \citep{Wild2010,Krishna2025}, suggests there is still sufficient material present at the beginning of the PSB phase, which could contribute to the obscuration of the central AGN. The aim of this work is to explore the possibility that massive, recently quenched galaxies may be harbouring dust-obscured AGN, by searching for their characteristic hot dust emission using the Mid-Infrared Instrument (MIRI) on \emph{JWST}.

The structure of the paper is as follows: In Section \ref{sec:Data} we describe the sample and the creation of the mid-IR catalogue, as well as the classification methods and extraction of key properties. In Section \ref{sec:Colour_sep} we present the mid-IR colours of PSBs and the F770W--F1800W/F444W--F770W colour-space which allows for the separation of star-forming and quiescent galaxies, as well as AGN sources. Section \ref{sec:AGN_models} explores the effect of AGN on the mid-IR colours of PSBs, providing an upper limit Eddington ratio for potentially obscured AGN. We discuss our findings in Section \ref{sec:Discussion} and the implications this has on our understanding of quenching pathways. We end with Section \ref{sec:Conclusions} where we summarize the work and give our conclusions.

We adopt the $\Lambda$CDM cosmology parameters in this work, with $\Omega_m=0.3$, $\Omega_\Lambda=0.7$, and $H_0=70$ km s$^{-1}$ Mpc$^{-1}$. All magnitudes stated are in the absolute bolometric (AB) system. Any uncertainties presented are given at the 1$\sigma$ level, unless otherwise stated.

\section{Data and Sample Selection}\label{sec:Data}
The data used in this work comes from the Public Release IMaging for Extragalactic Research (PRIMER) survey \citep{Dunlop2021,Donnan2024}. PRIMER is a large Cycle 1 \emph{JWST} Treasury programme that is a follow-up on the deeply-probed UKIDSS Ultra Deep Survey (UDS) field and the Cosmic Evolution Survey (COSMOS) field. The entire PRIMER survey was awarded 199.3 hours of observation and covers 234 sq. arcmin (125 sq. arcmin) of the UDS field and 143 sq. arcmin (110 sq. arcmin) in the COSMOS field with NIRCam (MIRI) imaging. It contains observations across seven NIRCam wide filters (F090W, F115W, F150W, F200W, F277W, F356W, F444W), one NIRCam medium filter (F410M), and two MIRI wide filters (F770W, F1800W). In addition, PRIMER is centred on the respective Cosmic Assembly Near-infrared Deep Extragalactic Legacy Survey (CANDELS) regions of both fields \citep{Grogin2011,Koekemoer2011}, allowing the addition of imaging from three blue \emph{HST}/ACS wide filters (F435W, F606W, F814W). In this work we focus on the observations from PRIMER-UDS.

\subsection{NIRCam and ACS photometric catalogue}
The PRIMER-UDS object detection and locations were based on the F200W band image, and extracted via \textsc{SExtractor} \citep{sextractor}. Aperture photometry was completed for all 11 NIRCam and ACS bands (with a diameter of $0.5^{\prime\prime}$), centred on the F200W detection locations. The F200W filter was chosen as the selection band to allow for a more direct comparison to the ground-based studies of the same field which were $K$-band selected \citep[e.g.][]{Wild2016,Wilkinson2021}. Stars and their diffraction spikes, as well as other artifacts, were masked out through a combination of the \texttt{CLASS\_STAR} parameter conditions and manual masking. Aperture photometry was corrected to the point-spread-function (PSF) on a band-by-band basis, or `corrected-to-total', which applies an additional normalization correction based on the F200W \texttt{MAG\_BEST} measurement (to obtain as close to the intrinsic flux measurements as possible, and to give valid stellar mass measurements). The full extraction method for these bands (as well as the sample classification and redshift fits), is described in de Lisle (\emph{submitted}), with a brief summary of the process presented here. 

The photometric redshifts of the sample galaxies were determined using the \textsc{eazy} software \citep{EAZY}. These values were validated using $\sim800$ high-quality spectra. The spectra were observed through a combination of the UDSz project \citep{Bradshaw2013,McLure2013}, the VANDELS project \citep{McLure2018,Pentericci2018}, and the EXCELS project \citep{Carnall2024}. The agreement distribution between the redshift measurements was $\frac{\Delta z}{1+z}\approx0.019$ over a redshift range of $0<z<5$, with an outlier amount of $\sim4\%$ (where outliers are defined as $\frac{\Delta z}{1+z}>0.15$). The complete photometric catalogue across $0<z<5$ contains 110 658 galaxies, which is cut down to 36 640 after applying a magnitude constraint of F200W $<26.5$.

\subsection{Super-colour classification}\label{sec:SC_classification}
The galaxies were classified through the SC method mentioned in Section \ref{sec:Intro}, using the ACS and NIRCam band photometry for the analysis. This approach uses PCA to identify key photometric trends that capture the maximal variation between galaxy SEDs (i.e. the method identifies the prominent eigenvectors of the population). The normalized SED of any galaxy, $f_\lambda/n$, can be described through a linear combination of a mean SED, $m_\lambda$, and a series of $p$ principal eigenvectors $e_{i\lambda}$ (or rather, `eigenspectra')
\begin{equation}
    \frac{f_\lambda}{n}=m_\lambda+\sum^p_{i=1}a_ie_{i\lambda}\:,
\end{equation}
where the amplitudes, $a_i$, of each eigenspectrum are the respective SC values for each galaxy. With the use of only three SCs the sample galaxy population can be classified into star-forming, quiescent, PSB, and dusty \citep[see][]{Wild2014}, similar to a conventional colour-colour space.
While the SCs do not directly correspond to individual properties, the first is correlated with the overall colour and mean stellar age of the galaxy, the second is correlated with the fraction of recently formed stars and thus the strength of the 4000\,\AA/Balmer break, and the third helps to further characterize the shape of the break and break metallicity degeneracies. The eigenbasis where the eigenspectra are constructed is based on a library of tens of thousands of `stochastic burst' model SEDs that are generated from the stellar population synthesis models of \citet{BruzalCharlot2003}, and with a \citet{Chabrier2003} initial mass function. During the SC analysis, additional physical properties are also fit for each galaxy, including stellar mass, age, star formation rate, and burst fraction. This is achieved in a Bayesian manner, building off of the same model SEDs that created the SC eigenbasis. Posterior probability density functions (PDFs) of each property are created from models with corresponding SC values, with the median value of each posterior PDF given as the property value for the corresponding real galaxy (with the errors corresponding to the 1$\sigma$ range values of the matching distribution). Additional specifics to the PCA method can be found in \citet{Wild2014}, with further specifics to the method applied to the PRIMER-UDS sample found in de Lisle (\emph{submitted}).

\begin{figure}
\centering
\includegraphics[width=0.99\linewidth]{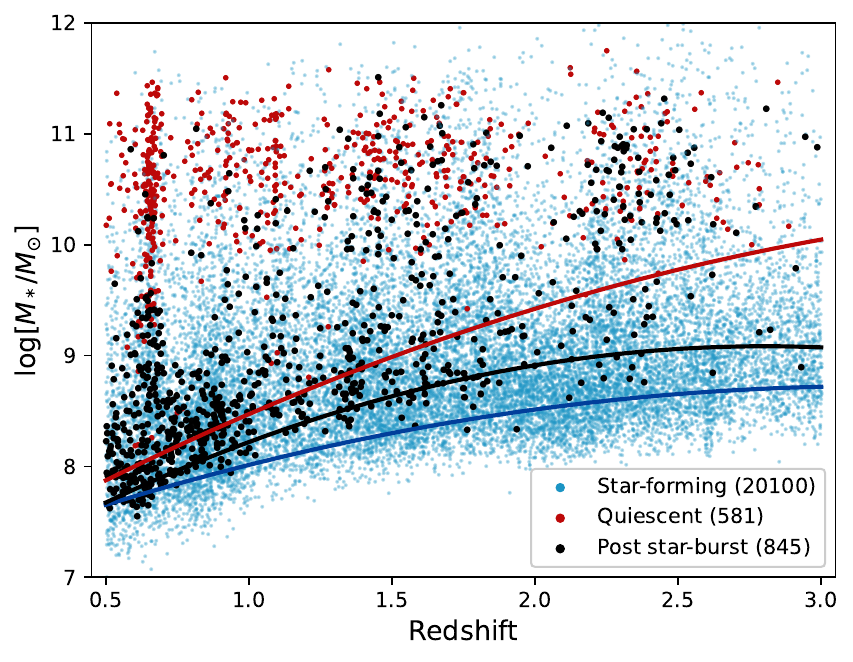}
\caption{Stellar mass of the sample galaxies against their photometric redshifts (or spectroscopic if available). The sample is selected from the F200W NIRCam filter image, with a magnitude completeness of $\text{mag}<26.5$ for $0.5^{\prime\prime}$ apertures. Overlaid are the $90\%$ stellar mass completeness curves of the star-forming, quiescent, and PSB populations, following the procedure in \citet{Pozzetti2010}.}
\label{fig:completeness}
\end{figure}

The SC classifications were based on the $0.5^{\prime\prime}$ diameter aperture measurements of each object (`corrected-to-total'). While the depth of the F200W image is $\simeq27.5$ (5$\sigma$; $0.5^{\prime\prime}$ apertures), a conservative magnitude limit of 26.5 was instead selected for the sample due to depth variation across the image, and to ensure sufficiently high S/N levels for robust SC classifications. The combined data across the ACS and NIRCam filters (0.4--4.4 $\mu$m) allows the overall sample to be classified from $0.5<z<3.0$ (which cuts the sample down to 29 970 galaxies), and down to at least $\log_{10}\left({M_*/\textrm{M}_\odot}\right)=9.0$ for PSBs across this redshift range. This makes it possible to explore the entirety of the PSB population, including both the distinct high-mass and low-mass PSB sub-populations. The $90\%$ stellar mass completeness curves for the sample are shown in Figure \ref{fig:completeness}, and follow the method of \citet{Pozzetti2010}. The sample has 21526 (18033) classified galaxies, with 20100 (16700) star-forming, 581 (580) quiescent, and 845 (753) PSBs, across the entire mass range (above their respective 90\% completeness limits). The sample is constrained to the galaxies above the 90\% completeness limits. It should also be noted that in this work we treat both the star-forming and dusty populations as star-forming.

As the SC classification of galaxies are determined by their photometry, increased photometric errors can lead to higher chances of misclassifications, particularly at low stellar mass.  Misclassifications of star-forming galaxies specifically are likely to dominate, simply due to their population size. In de Lisle (\emph{submitted}), the star-forming to PSB contamination rate is calculated, and estimated to be $<20\%$ at the mass completeness limit of our PSB sample. We note however, that this value is specifically at the lower mass limit and potential contamination drops dramatically at higher stellar  mass, with contamination rates at $\log_{10}\left({M_*/\textrm{M}_\odot}\right)=10$ being $\sim3\%$, for example.

Another source of misclassification can occur with true PSBs being misclassified as star-forming. This can occur due to the presence of blue light from  \emph{unobscured} AGN light which, if dominant, can alter the global SED and lead to a star-forming classification. The effect this has on SC classifications is explored in \citet{Almaini2025}, where they find PSBs with AGN reddened by dust with $A_V\sim1$ need to be accreting at $L_{AGN}/L_{Edd}>0.1$ to cause the misclassification. The likelihood of misclassification drops significantly with $A_V$ however, as PSBs hosting heavily obscured AGN ($A_V>2$) are expected to remain within the PSB classification region for 
$L_{AGN}/L_{Edd}\leq1$.  Since  this work is focused on the effects of highly obscured AGN ($A_V>\sim2$), 
we anticipate that the likelihood of supercolour misclassification due to contaminating AGN light is low for this population. This issue is discussed further in Section \ref{sec:Future_work}.

\subsection{MIRI photometric catalogue}
The photometric catalogue for the two MIRI bands was constructed in this work, using the v3.0 internal release PRIMER-UDS images that were reduced using the procedures in \citet{PerezGonzalez2024}. Fluxes were measured via forced photometry at positions defined by the F200W detection image, consistent with those used for the NIRCam/ACS photometry. Due to the size of the PSF in both MIRI bands, an aperture diameter of $1.0^{\prime\prime}$ was adopted for their photometry measurements. Using the calibration PSF images from the MIRI commissioning programme \citep[PID 1028, see][]{Dicken2024}, aperture corrections were applied to the photometry in each band. Alternative aperture corrections based on stacked stars from the PRIMER field were also explored, yielding MIR colours consistent with those from the calibrated corrections, with no impact on the results presented in Sections \ref{sec:Colour_sep} and \ref{sec:AGN_models}. All colours and SED analyses in this work use PSF-corrected MIRI, NIRCam, and ACS photometry.

The errors on the MIRI photometry followed the same procedure as the NIRCam/ACS, being calculated through a combination of intrinsic flux errors and background errors. A conservative absolute flux error was set as $5\%$ of the intrinsic flux (non-PSF-corrected). The background errors were calculated by sampling apertures placed at regular intervals across the entire field, which had been turned into a pure background image through the removal of dilated masks of all sources, artifacts, and image borders. The variance ($1\sigma$) of all aperture fluxes within a $25^{\prime\prime}$ radius circle around each object is taken as their background error. The total error is the flux error and the background error added in quadrature.

While all MIRI measurements are forced photometry based on F200W detections (which are >5$\sigma$ above the background), the photometric values for each object are only taken into account if they are >1$\sigma$ above the respective MIRI filter background (the non PSF-corrected $1.0^{\prime\prime}$ $5\sigma$ magnitude depths for F770W and F1800W are 25.12 and 22.83, respectively). While this further constrains our sample, it ensures that the MIRI fluxes for all objects are more robust against contamination from background fluctuations.\footnote{We verify that this constraint does not bias the massive galaxy populations, with at most only a single galaxy being removed from each group (star-forming, quiescent, and PSB). For the low mass populations it removes a small population ($\sim10\%$) of unconstrained sources that otherwise increase scatter. The results presented in Section \ref{sec:AGN_models} are robust to this cut.} Due to the smaller area of the MIRI field, our sample of galaxies with any MIR coverage is 10996 star-forming, 368 quiescent, and 464 PSBs. After applying all the additional cuts described in this section, our final sample contains 3519 star-forming, 226 quiescent, and 120 PSBs across $0.5<z<3.0$.

\section{The Mid-IR colours of post-starbursts}\label{sec:Colour_sep}
\begin{figure*}
\centering
\includegraphics[width=0.99\linewidth]{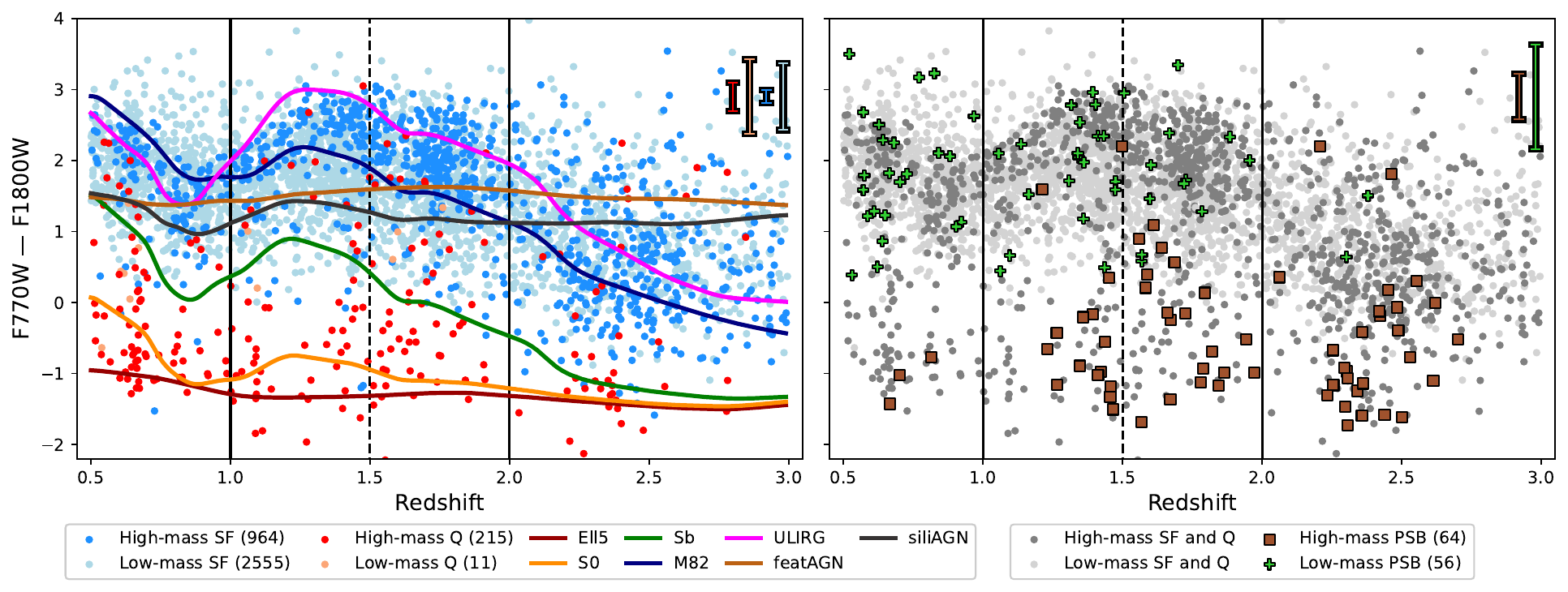}
\caption{\emph{Left}: Comparison of the F770W--F1800W colour of all quiescent and star-forming galaxies (as classified by the PCA technique), against their redshift. Each population is split at the stellar mass limit of $10^{10}\textrm{\;M}_\odot$, with the red and blue points showing the high-mass quiescent and high-mass star-forming population, respectively, and the light-red and light-blue points showing the respective low-mass populations. Also shown are the equivalent colour and redshift trends of five galaxy spectra from from the SWIRE library \citep{Polletta2007}, and two composite AGN templates from \citep{Kirkpatrick2012}. Overplotted are two solid black lines indicating the redshift range of $1<z<2$ which we focus on, due to the enhanced separation of the two populations (and the dashed black line indicating the bin split between them). At the top right of the plot the median colour errors for the quiescent and star-forming populations are shown, with their colours corresponding to their relevant population.  \emph{Right}: Same structure as the left panel, but the star-forming and quiescent galaxies are combined (with the grey and light-grey points showing the respective high- and low-mass populations) to give emphasis to the added PSB population. Massive PSBs are denoted as brown squares and low-mass PSBs are denoted as green crosses. The median colour errors for each population are shown at the top right of the plot. There is a distinct separation between the colours of these two PSB populations, with the low-mass PSBs containing an excess of MIR emission.}
\label{fig:miri_colour}
\end{figure*}

As the presence of an obscured AGN would result in excess emission in the MIR, it would be expected that this effect would be evident in the broad-band MIR colours of the galaxies \citep{Lacy2004,Nenkova2008}. The left panel of Figure \ref{fig:miri_colour} shows the F770W--F1800W colour of the star-forming and quiescent galaxies against their redshifts, for the entire sample range ($0.5<z<3.0$). In lieu of explicitly showing the colour errors on each data point, each population is split on the stellar mass delineation of $10^{10}\textrm{\;M}_\odot$, with the less massive population shown in lighter shades of blue and red, allowing the general redshift-colour trends of the more constrained (massive) population to be more easily seen (it should be noted however that the median colour error for each population is shown at the top right of each panel). Regardless of this mass cut, the overall star-forming population consistently has increased (redder) colours when compared to the quiescent population. This indicates that star-forming galaxies have elevated emission in the MIR, as expected, since F770W--F1800W probes the hot dust component, which can be heated by star formation. This follows well with the results from literature, where the use of some MIR colour equivalents have been used to separate local universe star-forming and quiescent galaxies, with the colour often being dubbed a `dust-to-star ratio' \citep{Li2007,Wilman2008}. It is important to stress that the SC classifications were evaluated \emph{without} the MIRI photometry, which adds emphasis that the F770W--F1800W colour provides a very effective technique for separating passive and star-forming galaxies.

\begin{figure}
\centering
\includegraphics[width=0.99\linewidth]{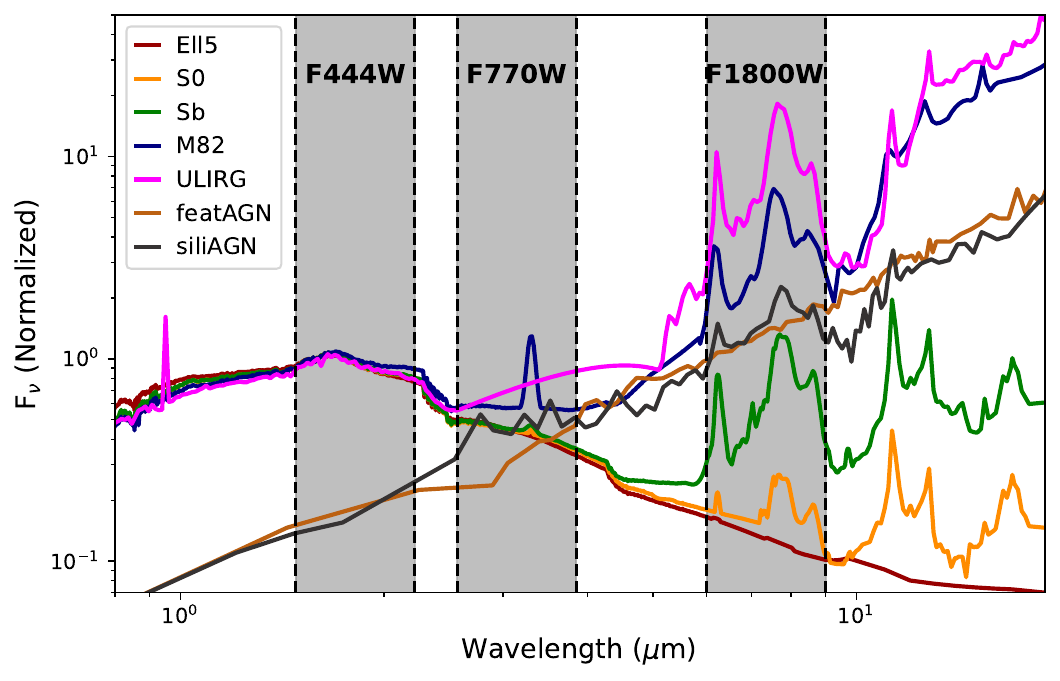}
\caption{Visualization of physical spectral features being probed by the F444W, F770W, and F1800W filters at $1<z<2$. Vertical grey columns correspond to the restframe of the filter effective wavelength probed across our redshift range ($1<z<2$). The galaxy SED templates from \citet{Polletta2007} (\emph{Ell5}, \emph{S0}, \emph{Sb}, \emph{M82}, \emph{ULIRG}) are normalized to 1.6 $\mu$m, and the AGN templates from \citep{Kirkpatrick2012} (\emph{featAGN}, \emph{siliAGN}) are normalized to 6 $\mu$m.}
\label{fig:colour_expl}
\end{figure}

While the trend of the quiescent population in Figure \ref{fig:miri_colour} is relatively flat, there is visible structure in the star-forming population through redshift, especially at $z<2$. In order to validate that these trends are physical, example spectra of observed galaxies \citep[from the SWIRE library of][]{Polletta2007} are taken and convolved through the related MIRI filters, to generate their F770W--F1800W colours through the redshift range of our sample. The spectral template colours cover the range of the observed colours well, and also exhibit the same broad redshift trends seen especially in the massive populations, with the star-forming and starburst templates (\emph{M82} and \emph{ULIRG} in the Figure \ref{fig:miri_colour} legend, respectively) following the range of the massive star-forming population, the elliptical and lenticular templates (\emph{Ell5} and \emph{S0}) consistent with the range of the majority of the massive quiescent population, and the intermediate spiral (\emph{Sb}) template residing in an intermediate location. In addition, two composite MIR-selected AGN SEDs from \citet{Kirkpatrick2012} (created from AGN with silicate features (\emph{siliAGN}), and those that are featureless (\emph{featAGN})) are plotted, showing elevated yet consistent colour values, which themselves are generally comparable with the star-forming population throughout the entire redshift range. In the MIR, the SED shapes of Type I and Type II AGN are relatively consistent, as dust attenuation primarily affect UV and optical wavelengths. The SWIRE QSO templates were omitted from Figure \ref{fig:colour_expl} due to their lower spectral resolution in this regime, however, their MIR colours and trends match those of the \emph{featAGN} and \emph{siliAGN} templates, as expected.

The right panel of Figure \ref{fig:miri_colour} is of the same form as the left, however the star-forming and quiescent populations have been combined and the PSB population is now shown. The PSBs are separated by the same mass delineation as the star-forming and quiescent populations. The massive PSBs are denoted as brown squares, while the low-mass PSBs are denoted as green crosses. It can be seen that the high-mass PSBs are generally consistent with the quiescent colour values (as well as more predominant in the high-\emph{z} regime), while the low-mass PSBs are generally consistent with the star-forming colour values (and are the dominant PSB population at low-z). This further confirms results from previous works that show a distinct bimodality within the PSB population \citep{Wild2016,Maltby2018}, and will be discussed further in subsequent sections.

Both our data and spectral model templates show a clear separation in F770W--F1800W between star-forming and quiescent galaxies, particularly at $z<2$. Since AGN templates also exhibit a pronounced excess in this colour, constraining our sample to $z<2$ maximizes the ability to identify galaxies with a MIR excess. We split the remaining sample into three redshift bins, each with a bin size of $\Delta z=0.5$ (as indicated by the vertical black lines in Figure \ref{fig:miri_colour}). To preserve the clarity of our results, we omit the formal presentation of the lowest redshift bin ($0.5<z<1$). While the trends from this regime are consistent to those in the two higher redshift bins, the low sample statistics for the high-mass PSB population (three galaxies) results in redundant visual clutter. As such, we relegate the related plots from the $0.5<z<1$ bin to Appendix \ref{sec:Appendix}, and instead focus the discussion in the main work to the statistically significant regime.

In order to understand the colour trends further, we explore what spectral features are being probed by this redshift range. Figure \ref{fig:colour_expl} shows a zoom-in on the same spectral templates used in Figure \ref{fig:miri_colour}, expressed in $F_\nu$ to allow a direct comparison with the colours. The galaxy templates have been normalized to 1.6 $\mu$m to give a better comparison of their spectral shapes, while the AGN templates are normalized to 6 $\mu$m, to give an understanding of how they affect the overall galaxy colours when present at a non-dominating level. Three vertical grey columns are shown, corresponding to the range of the effective wavelengths that each filter probes across the $1<z<2$ range. It can be seen that while there is relatively small variation in the spectra within the region probed by F770W, there is significant variation across the samples in F1800W, with the overall flux corresponding with the level of star formation in each galaxy. The features picked up by F1800W include the 6.2 $\mu$m, 7.7 $\mu$m, and the 8.6 $\mu$m polycyclic aromatic hydrocarbon (PAH) features, which are caused by the vibrational fluorescence of PAHs excited by FUV photons \citep{Peeters2002,Tielens2008}. These features (as well as the increase in the underlying continuum) are therefore enhanced in the HII regions around massive stars, and as such their strengths act as a good SFR tracer (again as can be easily seen through the strong gradient of the templates) \citep{Calzetti2007,Smith2007}. It follows then why the star-forming galaxies have an enhanced F770W--F1800W colour. We can also see that the power-law structure of the AGN templates also lead to consistently enhanced colours, regardless of redshift, as shown in their curves in Figure \ref{fig:miri_colour}.

It is evident that, while effective at identifying MIR emission as a whole, the F770W--F1800W colour cannot distinguish the source of the emission (i.e. star formation or AGN) on its own. The F444W filter, which probes around the 2 $\mu$m restframe range (the first grey column in Figure \ref{fig:colour_expl}), can help break this degeneracy. While all the galaxy templates have blue F444W--F770W colours (with the star-forming population being slightly elevated due to the 3.3 $\mu$m feature), the AGN SED would result in significantly redder colours. As the AGN component grows in strength relative to the stellar component, the F444W--F770W colour would be elevated, even if the F770W--F1800W is consistent with other star-forming galaxies.

\begin{figure*}
\centering
\includegraphics[width=0.99\linewidth]{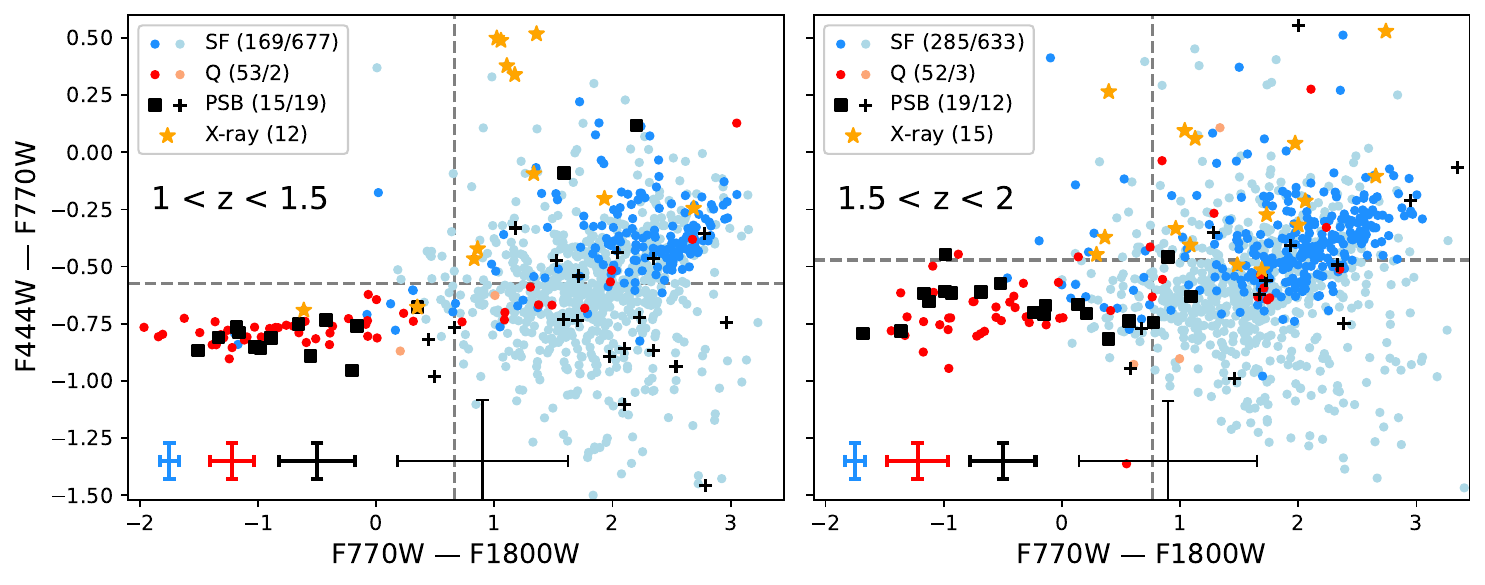}
\caption{Comparison of the F770W--F1800W colour and the F444W--F770W colour within the $1<z<1.5$ (\emph{left}) and $1.5<z<2$ (\emph{right}) redshift bins for the star-forming, quiescent, and PSB populations. Just like Figure \ref{fig:miri_colour} the populations are separated by a mass split at $10^{10}\textrm{\;M}_\odot$, with the black squares being the massive PSBs and the black crosses being the low-mass PSBs. The vertical and horizontal grey dashed lines indicate the respective separation location of the populations in F770W--F1800W and F444W--F770W (both based on the KDE minima of each of the massive star-forming and quiescent colour distributions), with galaxies exhibiting a colour redder (right) of the vertical cut considered to have a \emph{MIR excess}, and those with a colour bluer (left) of the vertical cut considered to have a \emph{MIR non-excess}. Galaxies that have X-ray emission detected by \emph{Chandra} imaging are denoted by yellow stars. At the bottom, median 1$\sigma$ colour errors of the populations are shown, corresponding to the massive star-forming, massive quiescent, massive PSB, and low-mass PSB, from left to right. Note: The values for each population given in the legend correspond only to the number of objects within the given colour range of the plots. The few objects just outside this range are almost exclusively low-mass SF galaxies, due to their elevated colour errors.}
\label{fig:colourcolour}
\end{figure*}

The use of both colours in a colour-colour space is shown in Figure \ref{fig:colourcolour}, with each plot corresponding to each of the redshift bins. Since F444W is the reddest NIRCam band we use an $1.0^{\prime\prime}$ aperture for consistency with the MIRI filters (and applying its PSF correction). Each of these plots use the same colour scheme as in Figure \ref{fig:miri_colour}, with the high-mass star-forming and quiescent galaxies shown in blue and red, respectively, with the lighter variant of both corresponding to their low-mass counterparts. Both plots show the star-forming and quiescent populations as significantly separated (with the massive star-forming population elevated in both colours), especially with the more constrained massive population (as indicated by the median errors shown at the bottom of the plot). This trend is consistent to that shown by the equivalent colour-colour space using the first three \emph{WISE} bands \citep{Wright2010,Alatalo2017,Cluver2017}.

To emphasize the distinction and colour separation of the populations, grey dashed lines are placed at the local minima of the combined kernel density estimates of the 1D colour distributions of the massive quiescent and star-forming populations, with the horizontal and vertical lines denoting the F444W--F770W and F770W--F1800W separation, respectively. The PSBs are also shown in these plots, again separated by the same stellar mass threshold (squares are massive, crosses are low-mass). It is evident that the vast majority of the massive PSBs sit left of the vertical line and below the horizontal line, thus exhibiting mid-IR colours entirely consistent with the quiescent population (i.e. containing very little hot dust). The low-mass PSBs on the other hand are consistent with the star-forming population, barring a larger spread due to their increased errors.

As previously discussed, the F444W--F770W colour is helpful for identifying significant AGN components. All objects in Figure \ref{fig:colourcolour} with a \emph{Securely} matched Chandra X-ray component \citep[following the definition in][]{Almaini2025} have a yellow star overlaid. As expected from the templates, these objects reside in a F770W--F1800W range consistent with the star-forming population, but have noticeably enhanced F444W--F770W colour values. The sources they are associated with are almost entirely classified as star-forming (21 out of 27 sources across both redshift bins), with five associated with quiescents and only one associated with a massive PSB. Again, this follows the findings from \citet{Almaini2025}, where massive PSBs exhibit suppressed X-ray emission. The results from this colour space give strong confirmation that not only do massive PSBs lack the MIR and NIR excess that is expressed by observed AGN, but also the excess that is expressed by star formation, with the massive PSBs being entirely consistent with the quiescent population. 

\section{Do post-starbursts contain hidden AGN?}\label{sec:AGN_models}
In order to validate and quantify the lack of obscured AGN within the PSB population, we turn to spectral models to see how the presence of an AGN would affect the broad-band colours of the PSB population. Using templates from X-\textsc{cigale} \citep{Yang2022} and implementing their SKIRTOR torus models \citep{Stalevski2012,Stalevski2016}, AGN SEDs at a variety of viewing angles were constructed, to investigate the influence of both Type I and Type II AGN. As noted in Section \ref{sec:Colour_sep}, dust attenuation primarily affects the UV and optical regimes, however, it does extend into the NIR and slightly impacts the restframe wavelengths probed by our high-redshift bins (but this effect diminishes with decreasing redshift, as shown in Appendix \ref{sec:Appendix}). Consequently, we include the full range of attenuation in our analysis, paying particular attention to the heavily obscured templates.

\begin{figure*}
\centering
\includegraphics[width=0.99\linewidth]{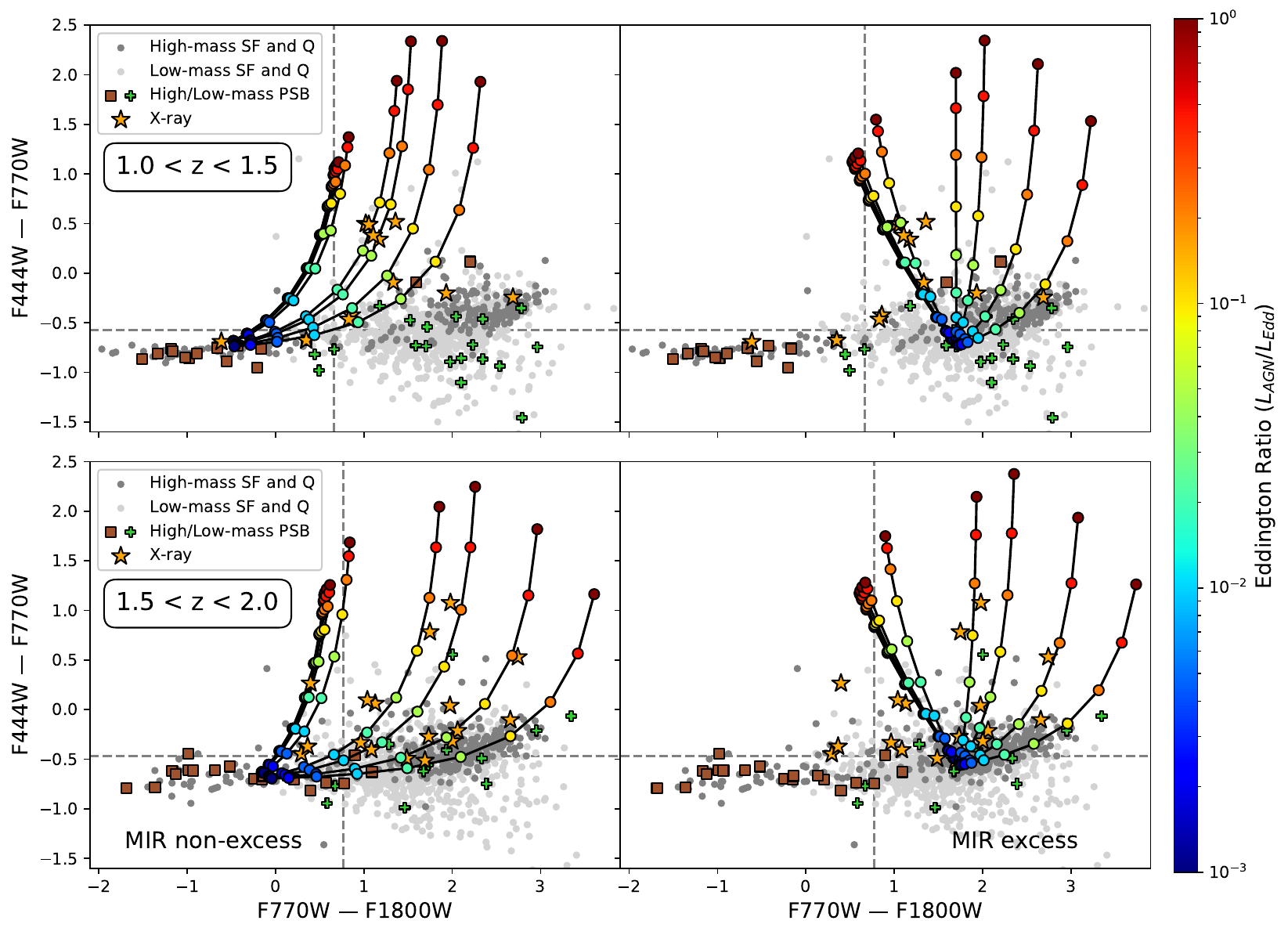}
\caption{Colour-colour diagram of galaxies, matching the form shown in Figure \ref{fig:colourcolour}, but with tracks overlaid indicating the influence of adding an AGN SED template to an individual PSB. Each row corresponds to a different redshift bin, indicated within the figure. The left panel in each row shows the tracks for the PSB with colour values closest to the median colours of the MIR non-excess population (left of the vertical dashed line). The right panel in each row shows the tracks for the PSB with colour values closest to the median colours of the MIR excess population (right of the vertical dashed line). Each track indicates a different viewing angle, beginning from $0^\circ$ (face-on) on the left and increasing by increments of $10^\circ$ up to an angle of $90^\circ$ (edge-on) with the right-most track. The point position on each track depends on the Eddington ratio applied, with specific intervals from $0.1\%$ to $100\%$ being shown and coloured according to the scale on the right-hand side of the figure.}
\label{fig:agn_tracks}
\end{figure*}

The AGN SEDs were related to the corresponding Eddington luminosity of each galaxy through the relations $M_\textrm{BH}=1.5\times10^{-3}M_*$ \citep{McLure2002,Haring2004} and $L_\textrm{Edd}=1.26\times10^{31}M_\textrm{BH} \textrm{\;J\:s}^{-1}$ \citep[following a similar prescription to the modelling in section 5 of][]{Almaini2025}, where $M_*$ is the stellar mass of the galaxy. The SEDs were normalized to bolometric using the median radio-quiet template from \citet{Elvis1994} (the normalization did not differ with the use of the radio-loud template), and scaled depending on the desired level of accretion (a factor from 0.1\% to 100$\%$ Eddington). They were then redshifted to match their host galaxy, and convolved through the desired NIRCam and MIRI filters. The resulting AGN fluxes were then added to the observed photometry of their host galaxy. Figure \ref{fig:agn_tracks} shows the AGN tracks of four example PSB galaxies, a MIR non-excess (\emph{left}) and a MIR excess (\emph{right}) for each redshift bin (\emph{top} = low redshift, \emph{bottom} = high redshift). Each example galaxy contains MIR colours that are representative of the median of each PSB population. Each of the tracks are a different viewing angle, increasing from face-on (0$^\circ$) on the left, to edge-on (90$^\circ$) on the right. The five lefthand-most tracks (the smallest angles) in all panels are extremely similar, due to the torus of the model only extending from $45^\circ \, \text{to} \, 90^\circ$, and as such the face-on templates ($0^\circ \, \text{to} \, 40^\circ$) do not change significantly. It can be seen that the tracks move away from the fiducial PSB example colours (especially in F444W--F770W) as the Eddington ratio increases, with the tracks approaching the general range where the X-ray sources dominate, regardless of viewing angle. The tracks for AGN at higher viewing angles more rapidly achieve elevated F770W--F1800W colours than the tracks of lower viewing angles.

From Figure \ref{fig:agn_tracks} we conclude that in order for the massive PSBs to be hiding highly accreting, obscured AGN, they would have to be residing in the top right of the overall colour-colour plot. However, the tracks also show that there is a smooth transition to this location, and the MIR colours of our observed massive PSBs are not inconsistent with hosting AGN at low Eddington ratios (e.g. $<1\%$). By applying these tracks to the entire population we can constrain an approximate upper limit value for AGN accretion within PSBs.

\begin{figure*}
\centering
\includegraphics[width=0.9\linewidth]{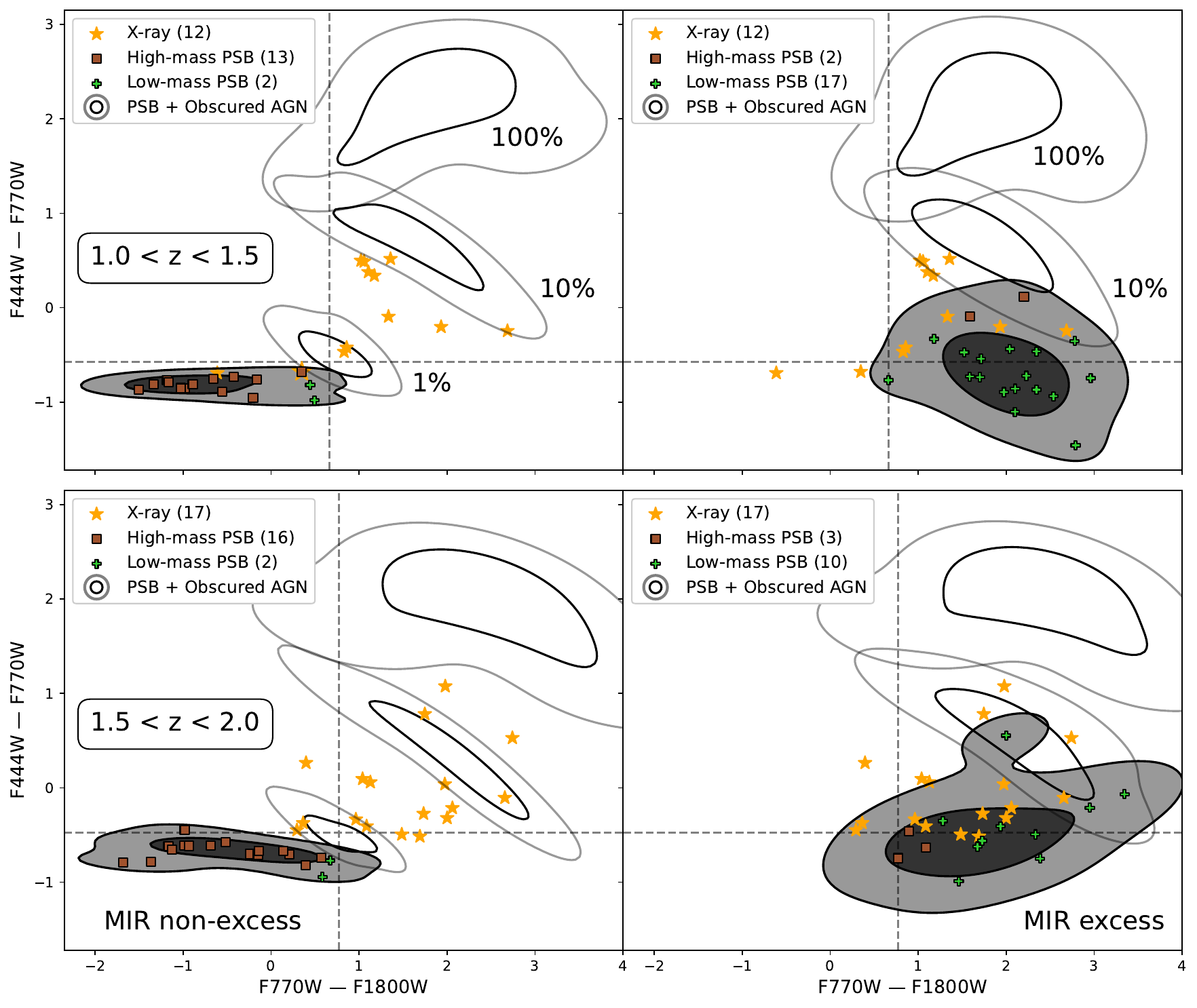}
\caption{Colour-colour locations and probability distributions of PSBs, and how their colours change with the addition of an AGN template accreting at various Eddington ratios. The left panels show the fiducial location of all MIR non-excess PSBs (F770W--F1800W values less than the vertical dashed line), and the right panels show the fiducial location of all MIR excess PSBs (with high-mass PSBs denoted as brown squares and low-mass as green crosses). Each row corresponds to a different redshift bin. In each panel the black and grey contours indicate the 1$\sigma$ and 2$\sigma$ probability distributions, respectively, with the filled black and grey contours indicating the same but for the fiducial PSB population. The clear contours in each panel indicate the colour location of PSBs when AGN with viewing angles of $50^\circ \, \text{to} \, 90^\circ$ are added, indicating the effect of an obscured AGN. For the MIR non-excess populations, contours indicating the addition of AGN accreting at $1\%$, $10\%$, and $100\%$ Eddington are shown, while for the MIR excess population only $10\%$ and $100\%$ are shown due to the similarity of the $1\%$ contours with the fiducial population. Additionally, the locations of all sources with secure X-ray detection are denoted by yellow stars, regardless of their SC classification.}
\label{fig:agn_probs}
\end{figure*}

Figure \ref{fig:agn_probs} shows how the combined colour space of all high-mass (and low-mass) PSBs change with the addition of an AGN accreting at a variety of Eddington ratios. To isolate the effect of obscured AGN, only AGN SEDs with viewing angles of $50^\circ \, \text{to} \, 90^\circ$ were added. In the left panels, which show the colour change for the MIR non-excess PSBs (effectively the massive PSBs), contours are given for Eddington ratios of $1\%$, $10\%$, and $100\%$. The contours were computed using KDE with a two-dimensional Gaussian kernel. The bandwidth of the kernels were scaled to a factor of 0.75 relative to the standard deviation of the data, to suppress small-scale Poisson noise and highlight the primary population structures. The different contours show that the observed PSB colours are inconsistent with the presence of AGN accreting at $10\%$ Eddington at a significance of $>8\sigma$ and $>5\sigma$ for the low- and high-redshift bins, respectively. The significance of the difference inevitably decreases at lower Eddington ratios. To obtain an approximate upper limit to the possible AGN contribution, we reduce the Eddington ratios in the models until the separation in colour can only be rejected at $95\%$ confidence. This occurs at very low Eddington ratios of $\sim0.5\%$ and $\sim1\%$, respectively. While the MIR non-excess population lack excess MIR emission by definition, the population spans a colour range wide enough to potentially include galaxies that contain non-negligible AGN contributions, thus biasing this upper limit value. Potential contamination levels are tested through comparison of the bluest PSB (which should be devoid of AGN contribution), the \emph{Ell5} quiescent spectral template (which contains no AGN contribution), and the PSB with the median colours (the same as those used in Figure \ref{fig:agn_tracks}). While these objects initially reside in well-separated locations along the MIR non-excess locus, with the addition of  AGN accreting at  $1\%$ Eddington their colours are very similar, due to the fact the AGN MIR emission quickly dominates, bringing all fiducial colours to a consistent location. Taking this into account, the upper limit of possible AGN contributions for the MIR non-excess population still holds at $\sim1\%$, and is likely a conservative limit for much of the population.

The right-hand panels of Figure \ref{fig:agn_probs} in each row show the same contour setup but for the MIR excess PSBs (effectively the low-mass PSB population). For this group we see the difference between the fiducial PSB colours and the addition of AGN is much less significant, and therefore we omit the $1\%$ Eddington contours due to their high similarity. For this population, the $10\%$ Eddington contours are at a $\sim2\sigma$ separation for both redshift bins, meaning an AGN within them would have to be very highly accreting in order to give them significantly different MIR colours. This upper limit indicates that with only these colours available, it is not possible to rule out that low-mass PSBs could be hiding obscured AGN.

\section{Discussion}\label{sec:Discussion}
\subsection{MIR excess (low-mass) PSBs}
Unlike the high-mass population, the low-mass PSBs show MIR colours broadly consistent with star-forming galaxies. While the results from Figure \ref{fig:agn_probs} indicate that obscured AGN in low-mass PSBs cannot be ruled out, both luminous AGN and star formation can heat up dust and cause a MIR excess, as discussed previously. Although the F770W--F1800W/F444W--F770W colour-space does well with separating out highly accreting AGN, there is still degeneracy between low-level accretion and star formation. The fact that as a population the low-mass PSBs are elevated in F770W--F1800W but show no significant excess in F444W--F770W (consistent with the range of the low-mass star-forming galaxies), gives credence to the more natural explanation that they contain residual star formation, rather than highly accreting AGN. 

\begin{figure}
\centering
\includegraphics[width=0.99\linewidth]{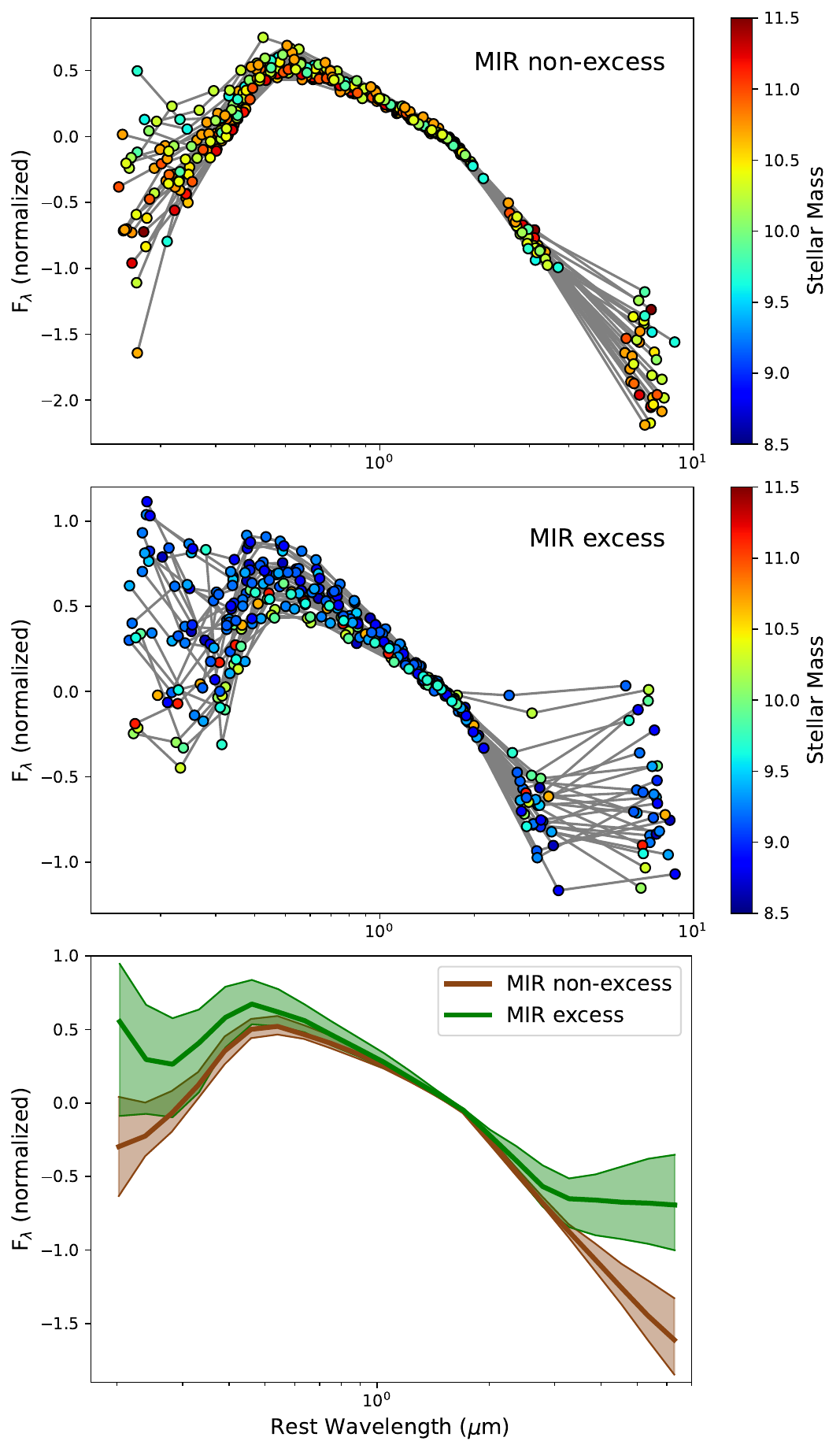}
\caption{SEDs of PSBs with MIR non-excess (\emph{top}) and with MIR excess (\emph{middle}). They have been shifted to their rest wavelength values based on their redshifts, and normalized to their interpolated 1.6 $\mu$m flux value. Each galaxy is coloured by their stellar mass values. The bottom panel compares the median SEDs of each population directly, again with the same normalization. The errors around the median is the standard deviation of the flux values across the population, added in quadrature to the median flux errors.}
\label{fig:psb_seds}
\end{figure}

The likelihood of residual star formation over AGN-heated dust is further enhanced when looking at the overall SEDs of each population. Figure \ref{fig:psb_seds} shows the photometry of MIR excess and MIR non-excess PSBs (top and middle panels, respectively). Each SED is shown at their restframe wavelengths, and they have all been normalized to their interpolated 1.6 $\mu$m flux value, for a better comparison of their shapes. The photometric points are coloured by their stellar mass, emphasising again how the MIR non-excess population contains almost exclusively high-mass galaxies, while the MIR excess population contains almost exclusively low-mass galaxies. In the bottom panel, the median SED of each population and their errors are shown. This panel makes clear the distinctness of their spectral shapes. Indeed by definition, the MIR excess population has an upturn in the MIR range, while the MIR non-excess has a smooth downturn. Interestingly, however, the MIR excess population also shows an upturn in the rest-frame UV regime below 0.3 $\mu$m (while the MIR non-excess has a downturn). This, combined with the enhanced peak at 0.4--0.5 $\mu$m (which is probing the A- and F- type star component), indicates that the low-mass PSBs are consistent with having residual star formation, or at least are closer to their starburst and quenching event than their high-mass counterparts.

Nevertheless, the quenching mechanism(s) for low-mass PSBs remains in question. They are distinct from their high-mass counterparts, with low-mass PSBs typically showing extended, disk-like morphologies, whereas high-mass PSBs appear as compact ellipticals \citep{Maltby2018,Cutler2024}. Furthermore, low-mass PSBs are preferentially found in higher density environments \citep{Wilkinson2021,Taylor2023}. The combination of distinct morphology and environment points to divergent quenching pathways for PSBs, where environmental processes play a dominant role for the low-mass population. The lack of AGN signatures would fit well into this interpretation, and even if present, AGN feedback is not thought capable of completely quenching these galaxies \citep{Tacchella2016}. Interestingly, our findings would also suggest that environmental processes do not completely quench low-mass PSBs either, but leave residual levels of star formation that we observe in our MIRI colours (and UV upturn).

A formal break of the degeneracy between star-formation and AGN heating would require SED fitting of the low-mass PSB photometry. However, fitting for dust-obscured AGN in this population is difficult as the low metallicity of dwarf galaxies means the typical PAH feature structure does not hold, as discussed in \citet{Rieke2025}. The lack of metals means stellar-heated dust can reach higher temperatures, both raising the MIR continuum and destroying the PAH features. The emission then more closely mimics the structure of AGN emission, allowing the signal to become easily misclassified, as shown in \citet{Hainline2016,Sturm2025}. Coupled with the MIR wavelength coverage we currently have in our sample, there would be little benefit to analysing the population with full SED fitting (although with extended coverage this may change, as we discuss in Section \ref{sec:Future_work}).

\subsection{MIR non-excess (high-mass) PSBs}
The primary result from this work shows that the vast majority of high-mass PSBs ($>10^{10}\textrm{\;M}_\odot$) show no excess MIR emission, indicating a lack of both obscured AGN and obscured star-formation. This implies that not only are their central SMBHs not continuously active throughout the entirety of the PSB phase, but moreover, highly accreting MIR AGN activity is fundamentally \emph{rare} in massive PSBs. This is consistent with the results from X-ray detections \citep[e.g.][]{French2023,Almaini2025} and radio detections \citep[e.g.][]{Luo2026,Patil2026}, however, it is now possible to eliminate the caveat that the lack of X-ray detections in PSBs could be due to them being heavily obscured as a population. Our results also allow us to constrain the strength of the MIR light that is present in PSBs, showing that as a population, their SMBHs are at accreting at $<1\%$ Eddington.

While these results allow for the possibility that high-mass PSBs are continuously in maintenance-mode ($<1\%$ Eddington) over their $\sim$1 Gyr visibility time, it is also possible that their AGN have short duty cycles, instead emitting in brief, stochastic, quasar mode ($>1\%$ Eddington) bursts rather than over a single monolithic period. Indeed, in our sample we find there are two massive PSBs that are elevated in both MIR colours (with both of these galaxies in the $1<z<1.5$ bin). Visual inspection of these objects shows that one of them is part of a complex, multi-galaxy merger, and thus likely has contaminated MIRI flux values, therefore we remove it from our analysis. The other elevated PSB is an isolated source with no immediate companions, with an upturn in both the MIR and the optical regime. Its F200W residual after the removal of a PSF convolved single S\'ersic model shows some extended structure which could be an indication of merger remnants, however, it is not significant enough to influence any definitive conclusions. This does mean however, that $\sim3\%$ ($1/33$) of our massive PSB sample have elevated MIR emission consistent with obscured AGN, which follows well with the equivalent $3\%$ found in both \citet{Meusinger2017} and \citet{Smercina2018} (both using low-\emph{z} WISE observations).

Within the massive PSB population there is also one source with a secure X-ray detection (again in the $1<z<1.5$ bin). Visual inspection shows it to also be isolated with no immediate companions, with no upturn in either the MIR or optical, and a F200W residual that suggests some excess internal structure, but no obvious merger remnants. A simple explanation for this object then could be a low-luminosity AGN. Indeed, based on its X-ray luminosity it has an Eddington ratio of $<0.01\%$, which would explain its lack of MIR excess. While the MIR colours at that accretion rate are consistent regardless of viewing angle, the presence of X-rays and the lack of MIR excess also points to it likely not being heavily obscured. In any case, this object gives the massive PSB population a fraction of $\sim3\%$ of X-ray detected AGN, which is consistent with the $\sim5\%$ value found in both \citet{French2023} and \citet{Almaini2025}. The low Eddington ratio also agrees with the results from \citet{Bugiani2025,Skarbinski2026}, which use line diagnostics to show that AGN in massive PSBs and quiescents at cosmic noon are, if present, relatively weak.

Our results imply that the duty cycle of AGN detected from X-rays or MIR are very similar. This is unsurprising as the MIR emission from AGN-heated dust should be coming from the inner torus region around the AGN, which has an extent of $\sim$ a few parsecs \citep{Jaffe2004,Tristram2007,Burtscher2013}, meaning that the light travel time would only be on the order of a few years. Therefore the duty cycles of the two should be very similar, assuming the AGN is active for timescales that are much longer than the travel time between the central SMBH and the torus. Indeed, \citet{French2023} calculates the duty cycle for the `on' and `fading' phases of AGN in PSBs, and finds those to be $\sim1\times10^4$ and $\sim1\times10^5$ years, respectively, which are much longer than the torus travel time. This connection is also shown explicitly in \citet{Sheng2017}, where the MIR light-curves of `changing-look AGN' match the trends of the optical component (with the offset of a few years), indicating that they can act as a dust echo, tracing variation in the AGN accretion rate.

\subsection{Future Work}\label{sec:Future_work}
Due to the size of the PSB sample in this work it is difficult to constrain the AGN duty cycle values to a high degree of certainty, or have a large population of massive MIR excess PSBs to analyse in more detail \citep[however it should be noted that our sample of 33 is equivalent to that in][]{Smercina2018}. Extending this analysis to include the equivalent imaging from the PRIMER-COSMOS field would help boost the sample by $\sim60\%$.

While the colour-space used in this work is highly effective at identifying MIR emission and highly accreting AGN, it would benefit from additional photometry. MINERVA \citep{Muzzin2025} is a follow-up survey on the PRIMER fields and will add imaging in the F1280W (12.8 $\mu$m) and F1500W (15 $\mu$m) bands (as well as eight NIRCam medium band filters). This, combined with our current MIRI imaging, would give continuous coverage around the majority of the PAH features, allowing the use of more optimized MIRI colour combinations to separate between AGN, composites, and star-forming galaxies \citep{Kirkpatrick2017,Chien2024,Kilerci2025,Vidal2026}. This would be especially effective for confirming whether the MIR excess in low-mass PSBs is due to AGN or residual star-formation.

The additional filters would also give us enough wavelength coverage to warrant SED fitting of our sample, to formally model the AGN component. One caveat within this analysis (as discussed in Section \ref{sec:SC_classification}), is that there are a potential number of PSBs that, due to the presence of unobscured AGN within them, have been misclassified as star-forming \citep[see][]{Almaini2025,Krishna2025}. If this is the case it could have implications on the true duty cycle values for PSBs. Therefore, being able to fit the entire sample would benefit our understanding of PSBs as a whole.

\section{Conclusions}\label{sec:Conclusions}
In this work we measure the MIR emission for all objects within the PRIMER-UDS field using the \emph{JWST}/MIRI F770W and F1800W bands. We create a colour-space (F770W--F1800W/F444W--F770W) to analyse and classify the MIR components of the galaxies, specifically to look for signals of obscured AGN within massive ($>10^{10}\textrm{\;M}_\odot$) PSBs. Our main findings are:
\begin{itemize}
    \item From the significant separation of massive PSBs from X-ray sources and the massive star-forming population, it is evident that most massive PSBs do not contain highly accreting obscured AGN, nor significant levels of obscured star formation. Instead, their MIR colours are entirely consistent with those of older passive galaxies.
    \item We find that $\sim3\%$ ($1/33$) of the massive PSBs have MIR excess consistent with the presence of obscured AGN. We also find that $\sim3\%$ ($1/33$) of massive PSBs have an X-ray detection. Overall, our results suggest that luminous AGN are rare in the PSB phase, consistent with the $\sim5\%$ duty cycle implied by previous studies at low- and high-redshift \citep[e.g.][]{French2023,Almaini2025}.
    \item We model how the presence of moderately- and highly-obscured AGN affect the MIR colours of PSBs and have shown that if present in the population they must have low (non-quasar) levels of accretion, on the order of $<1\%$ Eddington.
    \item We find that low-mass ($<10^{10}\textrm{\;M}_\odot$) PSBs have a MIR excess as well as a UV excess. While their MIR colours cannot rule out low-accretion AGN, their SED structure is indicative instead of residual star formation, indicating they are not completely quenched.
    \item These results add further weight to the growing body of evidence that suggests that low-mass PSBs follow quite different quenching pathways compared to massive PSBs.
\end{itemize}

In conclusion, our key finding is that the vast majority of massive PSBs show no evidence for highly accreting, obscured AGN. They also contain no evidence of obscured star-formation, or any significant residual star-formation. Low-mass PSBs remain a broadly unexplored population, with many questions about the distinctness of their quenching pathways. The addition of more fields (e.g. PRIMER-COSMOS), and addition wavelength coverage (e.g. MINERVA), should help increase the sample size and answer some of these questions.

\section*{Acknowledgements}
GH acknowledges the support of STFC grant ST/Y509437/1. OA acknowledges the support of STFC grant ST/X006581/1. EC acknowledges the support of a Royal Society Dorothy Hodgkin Fellowship Award DHF\textbackslash R\textbackslash 241008. KR gratefully acknowledges support from the NASA Astrophysics Data Analysis Program (ADAP) under grant 80NSSC23K0495. ET acknowledges support from a UKRI Frontier Research Guarantee Grant (PI Carnall; grant reference EP/Y037065/1). ACC thanks the Leverhulme Trust for their support via the Leverhulme Early Career Fellowship scheme. JSD and DJM acknowledge the support of the Royal Society, through the award of a Royal Society Research Professorship to JSD. PGP-G acknowledges support from grant PGC2018-093499-B-I00 funded by MCIN/AEI/10.13039/501100011033.

\emph{Software}: {\sc Astropy} \citep{astropy:2013,astropy:2018,astropy:2022}, {\sc EAZY} \citep{EAZY}, {\sc Matplotlib} \citep{Matplotlib:2007}, {\sc Numpy} \citep{Numpy2020}, {\sc Pandas} \citep{Pandas2010,Pandas2024}, {\sc Scipy} \citep{Scipy2020}, {\sc SExtractor} \citep{sextractor}.

\section*{Data Availability}

All \textit{JWST} and \textit{HST} data products are available via the Mikulski Archive for
Space Telescopes (\hyperlink{}{\textcolor{blue}{https://mast.stsci.edu}}).


\bibliographystyle{mnras}
\bibliography{main}

@ARTICLE{Almaini2025,
       author = {{Almaini}, Omar and {Wild}, Vivienne and {Maltby}, David and {Taylor}, Elizabeth and {Rowlands}, Kate and {de Lisle}, Thomas and {Alatalo}, Katherine and {Harrold}, Jimi and {Hewitt}, Guillaume and {Patil}, Pallavi and {Skarbinski}, Maya},
        title = "{No evidence for excess AGN activity in recently quenched massive galaxies at cosmic noon}",
      journal = {\mnras},
         year = 2025,
        month = jun,
       volume = {539},
       number = {4},
        pages = {3568-3581},
          doi = {10.1093/mnras/staf659},
archivePrefix = {arXiv},
       eprint = {2504.15342},
 primaryClass = {astro-ph.GA},
       adsurl = {https://ui.adsabs.harvard.edu/abs/2025MNRAS.539.3568A}
}

@ARTICLE{Strateva2001,
       author = {{Strateva}, Iskra and {Ivezi{\'c}}, {\v{Z}}eljko and {Knapp}, Gillian R. and {Narayanan}, Vijay K. and {Strauss}, Michael A. and {Gunn}, James E. and {Lupton}, Robert H. and {Schlegel}, David and {Bahcall}, Neta A. and {Brinkmann}, Jon and {Brunner}, Robert J. and {Budav{\'a}ri}, Tam{\'a}s and {Csabai}, Istv{\'a}n and {Castander}, Francisco Javier and {Doi}, Mamoru and {Fukugita}, Masataka and {Gy{\H{o}}ry}, Zsuzsanna and {Hamabe}, Masaru and {Hennessy}, Greg and {Ichikawa}, Takashi and {Kunszt}, Peter Z. and {Lamb}, Don Q. and {McKay}, Timothy A. and {Okamura}, Sadanori and {Racusin}, Judith and {Sekiguchi}, Maki and {Schneider}, Donald P. and {Shimasaku}, Kazuhiro and {York}, Donald},
        title = "{Color Separation of Galaxy Types in the Sloan Digital Sky Survey Imaging Data}",
      journal = {\aj},
         year = 2001,
        month = oct,
       volume = {122},
       number = {4},
        pages = {1861-1874},
          doi = {10.1086/323301},
archivePrefix = {arXiv},
       eprint = {astro-ph/0107201},
 primaryClass = {astro-ph},
       adsurl = {https://ui.adsabs.harvard.edu/abs/2001AJ....122.1861S}
}

@ARTICLE{Baldry2004,
       author = {{Baldry}, Ivan K. and {Glazebrook}, Karl and {Brinkmann}, Jon and {Ivezi{\'c}}, {\v{Z}}eljko and {Lupton}, Robert H. and {Nichol}, Robert C. and {Szalay}, Alexander S.},
        title = "{Quantifying the Bimodal Color-Magnitude Distribution of Galaxies}",
      journal = {\apj},
         year = 2004,
        month = jan,
       volume = {600},
       number = {2},
        pages = {681-694},
          doi = {10.1086/380092},
archivePrefix = {arXiv},
       eprint = {astro-ph/0309710},
 primaryClass = {astro-ph},
       adsurl = {https://ui.adsabs.harvard.edu/abs/2004ApJ...600..681B}
}

@ARTICLE{Elbaz2007,
       author = {{Elbaz}, D. and {Daddi}, E. and {Le Borgne}, D. and {Dickinson}, M. and {Alexander}, D.~M. and {Chary}, R. -R. and {Starck}, J. -L. and {Brandt}, W.~N. and {Kitzbichler}, M. and {MacDonald}, E. and {Nonino}, M. and {Popesso}, P. and {Stern}, D. and {Vanzella}, E.},
        title = "{The reversal of the star formation-density relation in the distant universe}",
      journal = {\aap},
         year = 2007,
        month = jun,
       volume = {468},
       number = {1},
        pages = {33-48},
          doi = {10.1051/0004-6361:20077525},
archivePrefix = {arXiv},
       eprint = {astro-ph/0703653},
 primaryClass = {astro-ph},
       adsurl = {https://ui.adsabs.harvard.edu/abs/2007A&A...468...33E}
}

@ARTICLE{McGee2011,
       author = {{McGee}, Sean L. and {Balogh}, Michael L. and {Wilman}, David J. and {Bower}, Richard G. and {Mulchaey}, John S. and {Parker}, Laura C. and {Oemler}, Augustus},
        title = "{The Dawn of the Red: star formation histories of group galaxies over the past 5 billion years}",
      journal = {\mnras},
         year = 2011,
        month = may,
       volume = {413},
       number = {2},
        pages = {996-1012},
          doi = {10.1111/j.1365-2966.2010.18189.x},
archivePrefix = {arXiv},
       eprint = {1012.2388},
 primaryClass = {astro-ph.CO},
       adsurl = {https://ui.adsabs.harvard.edu/abs/2011MNRAS.413..996M}
}

@ARTICLE{Driver2006,
       author = {{Driver}, S.~P. and {Allen}, P.~D. and {Graham}, Alister. W. and {Cameron}, E. and {Liske}, J. and {Ellis}, S.~C. and {Cross}, N.~J.~G. and {De Propris}, R. and {Phillipps}, S. and {Couch}, W.~J.},
        title = "{The Millennium Galaxy Catalogue: morphological classification and bimodality in the colour-concentration plane}",
      journal = {\mnras},
         year = 2006,
        month = may,
       volume = {368},
       number = {1},
        pages = {414-434},
          doi = {10.1111/j.1365-2966.2006.10126.x},
archivePrefix = {arXiv},
       eprint = {astro-ph/0602240},
 primaryClass = {astro-ph},
       adsurl = {https://ui.adsabs.harvard.edu/abs/2006MNRAS.368..414D}
}

@ARTICLE{Mignoli2009,
       author = {{Mignoli}, M. and {Zamorani}, G. and {Scodeggio}, M. and {Cimatti}, A. and {Halliday}, C. and {Lilly}, S.~J. and {Pozzetti}, L. and {Vergani}, D. and {Carollo}, C.~M. and {Contini}, T. and {Le F{\'e}vre}, O. and {Mainieri}, V. and {Renzini}, A. and {Bardelli}, S. and {Bolzonella}, M. and {Bongiorno}, A. and {Caputi}, K. and {Coppa}, G. and {Cucciati}, O. and {de La Torre}, S. and {de Ravel}, L. and {Franzetti}, P. and {Garilli}, B. and {Iovino}, A. and {Kampczyk}, P. and {Kneib}, J. -P. and {Knobel}, C. and {Kova{\v{c}}}, K. and {Lamareille}, F. and {Le Borgne}, J. -F. and {Le Brun}, V. and {Maier}, C. and {Pell{\`o}}, R. and {Peng}, Y. and {Perez Montero}, E. and {Ricciardelli}, E. and {Scarlata}, C. and {Silverman}, J.~D. and {Tanaka}, M. and {Tasca}, L. and {Tresse}, L. and {Zucca}, E. and {Abbas}, U. and {Bottini}, D. and {Capak}, P. and {Cappi}, A. and {Cassata}, P. and {Fumana}, M. and {Guzzo}, L. and {Leauthaud}, A. and {Maccagni}, D. and {Marinoni}, C. and {McCracken}, H.~J. and {Memeo}, P. and {Meneux}, B. and {Oesch}, P. and {Porciani}, C. and {Scaramella}, R. and {Scoville}, N.},
        title = "{The zCOSMOS redshift survey: the three-dimensional classification cube and bimodality in galaxy physical properties}",
      journal = {\aap},
         year = 2009,
        month = jan,
       volume = {493},
       number = {1},
        pages = {39-49},
          doi = {10.1051/0004-6361:200810520},
archivePrefix = {arXiv},
       eprint = {0810.2245},
 primaryClass = {astro-ph},
       adsurl = {https://ui.adsabs.harvard.edu/abs/2009A&A...493...39M}
}

@ARTICLE{Ilbert2010,
       author = {{Ilbert}, O. and {Salvato}, M. and {Le Floc'h}, E. and {Aussel}, H. and {Capak}, P. and {McCracken}, H.~J. and {Mobasher}, B. and {Kartaltepe}, J. and {Scoville}, N. and {Sanders}, D.~B. and {Arnouts}, S. and {Bundy}, K. and {Cassata}, P. and {Kneib}, J. -P. and {Koekemoer}, A. and {Le F{\`e}vre}, O. and {Lilly}, S. and {Surace}, J. and {Taniguchi}, Y. and {Tasca}, L. and {Thompson}, D. and {Tresse}, L. and {Zamojski}, M. and {Zamorani}, G. and {Zucca}, E.},
        title = "{Galaxy Stellar Mass Assembly Between 0.2 < z < 2 from the S-COSMOS Survey}",
      journal = {\apj},
         year = 2010,
        month = feb,
       volume = {709},
       number = {2},
        pages = {644-663},
          doi = {10.1088/0004-637X/709/2/644},
archivePrefix = {arXiv},
       eprint = {0903.0102},
 primaryClass = {astro-ph.CO},
       adsurl = {https://ui.adsabs.harvard.edu/abs/2010ApJ...709..644I}
}

@ARTICLE{Muzzin2013,
       author = {{Muzzin}, Adam and {Marchesini}, Danilo and {Stefanon}, Mauro and {Franx}, Marijn and {McCracken}, Henry J. and {Milvang-Jensen}, Bo and {Dunlop}, James S. and {Fynbo}, J.~P.~U. and {Brammer}, Gabriel and {Labb{\'e}}, Ivo and {van Dokkum}, Pieter G.},
        title = "{The Evolution of the Stellar Mass Functions of Star-forming and Quiescent Galaxies to z = 4 from the COSMOS/UltraVISTA Survey}",
      journal = {\apj},
         year = 2013,
        month = nov,
       volume = {777},
       number = {1},
          eid = {18},
        pages = {18},
          doi = {10.1088/0004-637X/777/1/18},
archivePrefix = {arXiv},
       eprint = {1303.4409},
 primaryClass = {astro-ph.CO},
       adsurl = {https://ui.adsabs.harvard.edu/abs/2013ApJ...777...18M}
}

@ARTICLE{McLeod2021,
       author = {{McLeod}, D.~J. and {McLure}, R.~J. and {Dunlop}, J.~S. and {Cullen}, F. and {Carnall}, A.~C. and {Duncan}, K.},
        title = "{The evolution of the galaxy stellar-mass function over the last 12 billion years from a combination of ground-based and HST surveys}",
      journal = {\mnras},
         year = 2021,
        month = may,
       volume = {503},
       number = {3},
        pages = {4413-4435},
          doi = {10.1093/mnras/stab731},
archivePrefix = {arXiv},
       eprint = {2009.03176},
 primaryClass = {astro-ph.GA},
       adsurl = {https://ui.adsabs.harvard.edu/abs/2021MNRAS.503.4413M}
}

@ARTICLE{Kauffmann2003,
       author = {{Kauffmann}, Guinevere and {Heckman}, Timothy M. and {White}, Simon D.~M. and {Charlot}, St{\'e}phane and {Tremonti}, Christy and {Peng}, Eric W. and {Seibert}, Mark and {Brinkmann}, Jon and {Nichol}, Robert C. and {SubbaRao}, Mark and {York}, Don},
        title = "{The dependence of star formation history and internal structure on stellar mass for {}10$^{5}$ low-redshift galaxies}",
      journal = {\mnras},
         year = 2003,
        month = may,
       volume = {341},
       number = {1},
        pages = {54-69},
          doi = {10.1046/j.1365-8711.2003.06292.x},
archivePrefix = {arXiv},
       eprint = {astro-ph/0205070},
 primaryClass = {astro-ph},
       adsurl = {https://ui.adsabs.harvard.edu/abs/2003MNRAS.341...54K}
}

@ARTICLE{Bundy2006,
       author = {{Bundy}, Kevin and {Ellis}, Richard S. and {Conselice}, Christopher J. and {Taylor}, James E. and {Cooper}, Michael C. and {Willmer}, Christopher N.~A. and {Weiner}, Benjamin J. and {Coil}, Alison L. and {Noeske}, Kai G. and {Eisenhardt}, Peter R.~M.},
        title = "{The Mass Assembly History of Field Galaxies: Detection of an Evolving Mass Limit for Star-Forming Galaxies}",
      journal = {\apj},
         year = 2006,
        month = nov,
       volume = {651},
       number = {1},
        pages = {120-141},
          doi = {10.1086/507456},
archivePrefix = {arXiv},
       eprint = {astro-ph/0512465},
 primaryClass = {astro-ph},
       adsurl = {https://ui.adsabs.harvard.edu/abs/2006ApJ...651..120B}
}

@ARTICLE{Ilbert2013,
       author = {{Ilbert}, O. and {McCracken}, H.~J. and {Le F{\`e}vre}, O. and {Capak}, P. and {Dunlop}, J. and {Karim}, A. and {Renzini}, M.~A. and {Caputi}, K. and {Boissier}, S. and {Arnouts}, S. and {Aussel}, H. and {Comparat}, J. and {Guo}, Q. and {Hudelot}, P. and {Kartaltepe}, J. and {Kneib}, J.~P. and {Krogager}, J.~K. and {Le Floc'h}, E. and {Lilly}, S. and {Mellier}, Y. and {Milvang-Jensen}, B. and {Moutard}, T. and {Onodera}, M. and {Richard}, J. and {Salvato}, M. and {Sanders}, D.~B. and {Scoville}, N. and {Silverman}, J.~D. and {Taniguchi}, Y. and {Tasca}, L. and {Thomas}, R. and {Toft}, S. and {Tresse}, L. and {Vergani}, D. and {Wolk}, M. and {Zirm}, A.},
        title = "{Mass assembly in quiescent and star-forming galaxies since z ≃ 4 from UltraVISTA}",
      journal = {\aap},
         year = 2013,
        month = aug,
       volume = {556},
          eid = {A55},
        pages = {A55},
          doi = {10.1051/0004-6361/201321100},
archivePrefix = {arXiv},
       eprint = {1301.3157},
 primaryClass = {astro-ph.CO},
       adsurl = {https://ui.adsabs.harvard.edu/abs/2013A&A...556A..55I}
}

@ARTICLE{Kauffmann2004,
       author = {{Kauffmann}, Guinevere and {White}, Simon D.~M. and {Heckman}, Timothy M. and {M{\'e}nard}, Brice and {Brinchmann}, Jarle and {Charlot}, St{\'e}phane and {Tremonti}, Christy and {Brinkmann}, Jon},
        title = "{The environmental dependence of the relations between stellar mass, structure, star formation and nuclear activity in galaxies}",
      journal = {\mnras},
         year = 2004,
        month = sep,
       volume = {353},
       number = {3},
        pages = {713-731},
          doi = {10.1111/j.1365-2966.2004.08117.x},
archivePrefix = {arXiv},
       eprint = {astro-ph/0402030},
 primaryClass = {astro-ph},
       adsurl = {https://ui.adsabs.harvard.edu/abs/2004MNRAS.353..713K}
}

@ARTICLE{Balogh2004,
       author = {{Balogh}, Michael L. and {Baldry}, Ivan K. and {Nichol}, Robert and {Miller}, Chris and {Bower}, Richard and {Glazebrook}, Karl},
        title = "{The Bimodal Galaxy Color Distribution: Dependence on Luminosity and Environment}",
      journal = {\apjl},
         year = 2004,
        month = nov,
       volume = {615},
       number = {2},
        pages = {L101-L104},
          doi = {10.1086/426079},
archivePrefix = {arXiv},
       eprint = {astro-ph/0406266},
 primaryClass = {astro-ph},
       adsurl = {https://ui.adsabs.harvard.edu/abs/2004ApJ...615L.101B}
}

@ARTICLE{Baldry2006,
       author = {{Baldry}, I.~K. and {Balogh}, M.~L. and {Bower}, R.~G. and {Glazebrook}, K. and {Nichol}, R.~C. and {Bamford}, S.~P. and {Budavari}, T.},
        title = "{Galaxy bimodality versus stellar mass and environment}",
      journal = {\mnras},
         year = 2006,
        month = dec,
       volume = {373},
       number = {2},
        pages = {469-483},
          doi = {10.1111/j.1365-2966.2006.11081.x},
archivePrefix = {arXiv},
       eprint = {astro-ph/0607648},
 primaryClass = {astro-ph},
       adsurl = {https://ui.adsabs.harvard.edu/abs/2006MNRAS.373..469B}
}

@ARTICLE{Muzzin2012,
       author = {{Muzzin}, Adam and {Wilson}, Gillian and {Yee}, H.~K.~C. and {Gilbank}, David and {Hoekstra}, Henk and {Demarco}, Ricardo and {Balogh}, Michael and {van Dokkum}, Pieter and {Franx}, Marijn and {Ellingson}, Erica and {Hicks}, Amalia and {Nantais}, Julie and {Noble}, Allison and {Lacy}, Mark and {Lidman}, Chris and {Rettura}, Alessandro and {Surace}, Jason and {Webb}, Tracy},
        title = "{The Gemini Cluster Astrophysics Spectroscopic Survey (GCLASS): The Role of Environment and Self-regulation in Galaxy Evolution at z \raisebox{-0.5ex}\textasciitilde 1}",
      journal = {\apj},
         year = 2012,
        month = feb,
       volume = {746},
       number = {2},
          eid = {188},
        pages = {188},
          doi = {10.1088/0004-637X/746/2/188},
archivePrefix = {arXiv},
       eprint = {1112.3655},
 primaryClass = {astro-ph.CO},
       adsurl = {https://ui.adsabs.harvard.edu/abs/2012ApJ...746..188M}
}

@ARTICLE{SilkRees1998,
       author = {{Silk}, Joseph and {Rees}, Martin J.},
        title = "{Quasars and galaxy formation}",
      journal = {\aap},
         year = 1998,
        month = mar,
       volume = {331},
        pages = {L1-L4},
          doi = {10.48550/arXiv.astro-ph/9801013},
archivePrefix = {arXiv},
       eprint = {astro-ph/9801013},
 primaryClass = {astro-ph},
       adsurl = {https://ui.adsabs.harvard.edu/abs/1998A&A...331L...1S}
}

@ARTICLE{Croton2006,
       author = {{Croton}, Darren J. and {Springel}, Volker and {White}, Simon D.~M. and {De Lucia}, G. and {Frenk}, C.~S. and {Gao}, L. and {Jenkins}, A. and {Kauffmann}, G. and {Navarro}, J.~F. and {Yoshida}, N.},
        title = "{The many lives of active galactic nuclei: cooling flows, black holes and the luminosities and colours of galaxies}",
      journal = {\mnras},
         year = 2006,
        month = jan,
       volume = {365},
       number = {1},
        pages = {11-28},
          doi = {10.1111/j.1365-2966.2005.09675.x},
archivePrefix = {arXiv},
       eprint = {astro-ph/0508046},
 primaryClass = {astro-ph},
       adsurl = {https://ui.adsabs.harvard.edu/abs/2006MNRAS.365...11C}
}

@ARTICLE{Fabian2012,
       author = {{Fabian}, A.~C.},
        title = "{Observational Evidence of Active Galactic Nuclei Feedback}",
      journal = {\araa},
         year = 2012,
        month = sep,
       volume = {50},
        pages = {455-489},
          doi = {10.1146/annurev-astro-081811-125521},
archivePrefix = {arXiv},
       eprint = {1204.4114},
 primaryClass = {astro-ph.CO},
       adsurl = {https://ui.adsabs.harvard.edu/abs/2012ARA&A..50..455F}
}

@ARTICLE{DekelSilk1986,
       author = {{Dekel}, A. and {Silk}, J.},
        title = "{The Origin of Dwarf Galaxies, Cold Dark Matter, and Biased Galaxy Formation}",
      journal = {\apj},
         year = 1986,
        month = apr,
       volume = {303},
        pages = {39},
          doi = {10.1086/164050},
       adsurl = {https://ui.adsabs.harvard.edu/abs/1986ApJ...303...39D}
}

@ARTICLE{Hopkins2012,
       author = {{Hopkins}, Philip F. and {Quataert}, Eliot and {Murray}, Norman},
        title = "{Stellar feedback in galaxies and the origin of galaxy-scale winds}",
      journal = {\mnras},
         year = 2012,
        month = apr,
       volume = {421},
       number = {4},
        pages = {3522-3537},
          doi = {10.1111/j.1365-2966.2012.20593.x},
archivePrefix = {arXiv},
       eprint = {1110.4638},
 primaryClass = {astro-ph.CO},
       adsurl = {https://ui.adsabs.harvard.edu/abs/2012MNRAS.421.3522H}
}

@ARTICLE{WhiteFrenk1991,
       author = {{White}, Simon D.~M. and {Frenk}, Carlos S.},
        title = "{Galaxy Formation through Hierarchical Clustering}",
      journal = {\apj},
         year = 1991,
        month = sep,
       volume = {379},
        pages = {52},
          doi = {10.1086/170483},
       adsurl = {https://ui.adsabs.harvard.edu/abs/1991ApJ...379...52W}
}

@ARTICLE{Moore1996,
       author = {{Moore}, Ben and {Katz}, Neal and {Lake}, George and {Dressler}, Alan and {Oemler}, Augustus},
        title = "{Galaxy harassment and the evolution of clusters of galaxies}",
      journal = {\nat},
         year = 1996,
        month = feb,
       volume = {379},
       number = {6566},
        pages = {613-616},
          doi = {10.1038/379613a0},
archivePrefix = {arXiv},
       eprint = {astro-ph/9510034},
 primaryClass = {astro-ph},
       adsurl = {https://ui.adsabs.harvard.edu/abs/1996Natur.379..613M}
}

@ARTICLE{Moore1998,
       author = {{Moore}, Ben and {Lake}, George and {Katz}, Neal},
        title = "{Morphological Transformation from Galaxy Harassment}",
      journal = {\apj},
         year = 1998,
        month = mar,
       volume = {495},
       number = {1},
        pages = {139-151},
          doi = {10.1086/305264},
archivePrefix = {arXiv},
       eprint = {astro-ph/9701211},
 primaryClass = {astro-ph},
       adsurl = {https://ui.adsabs.harvard.edu/abs/1998ApJ...495..139M}
}

@ARTICLE{Smith2015,
       author = {{Smith}, R. and {S{\'a}nchez-Janssen}, R. and {Beasley}, M.~A. and {Candlish}, G.~N. and {Gibson}, B.~K. and {Puzia}, T.~H. and {Janz}, J. and {Knebe}, A. and {Aguerri}, J.~A.~L. and {Lisker}, T. and {Hensler}, G. and {Fellhauer}, M. and {Ferrarese}, L. and {Yi}, S.~K.},
        title = "{The sensitivity of harassment to orbit: mass loss from early-type dwarfs in galaxy clusters}",
      journal = {\mnras},
         year = 2015,
        month = dec,
       volume = {454},
       number = {3},
        pages = {2502-2516},
          doi = {10.1093/mnras/stv2082},
archivePrefix = {arXiv},
       eprint = {1509.02537},
 primaryClass = {astro-ph.GA},
       adsurl = {https://ui.adsabs.harvard.edu/abs/2015MNRAS.454.2502S}
}

@ARTICLE{GunnGott1972,
       author = {{Gunn}, James E. and {Gott}, J. Richard, III},
        title = "{On the Infall of Matter Into Clusters of Galaxies and Some Effects on Their Evolution}",
      journal = {\apj},
         year = 1972,
        month = aug,
       volume = {176},
        pages = {1},
          doi = {10.1086/151605},
       adsurl = {https://ui.adsabs.harvard.edu/abs/1972ApJ...176....1G}
}

@ARTICLE{Larson1980,
       author = {{Larson}, R.~B. and {Tinsley}, B.~M. and {Caldwell}, C.~N.},
        title = "{The evolution of disk galaxies and the origin of S0 galaxies}",
      journal = {\apj},
         year = 1980,
        month = may,
       volume = {237},
        pages = {692-707},
          doi = {10.1086/157917},
       adsurl = {https://ui.adsabs.harvard.edu/abs/1980ApJ...237..692L}
}

@ARTICLE{Balogh2000,
       author = {{Balogh}, Michael L. and {Navarro}, Julio F. and {Morris}, Simon L.},
        title = "{The Origin of Star Formation Gradients in Rich Galaxy Clusters}",
      journal = {\apj},
         year = 2000,
        month = sep,
       volume = {540},
       number = {1},
        pages = {113-121},
          doi = {10.1086/309323},
archivePrefix = {arXiv},
       eprint = {astro-ph/0004078},
 primaryClass = {astro-ph},
       adsurl = {https://ui.adsabs.harvard.edu/abs/2000ApJ...540..113B}
}

@ARTICLE{Peng2015,
       author = {{Peng}, Y. and {Maiolino}, R. and {Cochrane}, R.},
        title = "{Strangulation as the primary mechanism for shutting down star formation in galaxies}",
      journal = {\nat},
         year = 2015,
        month = may,
       volume = {521},
       number = {7551},
        pages = {192-195},
          doi = {10.1038/nature14439},
archivePrefix = {arXiv},
       eprint = {1505.03143},
 primaryClass = {astro-ph.GA},
       adsurl = {https://ui.adsabs.harvard.edu/abs/2015Natur.521..192P}
}

@ARTICLE{Dressler1983,
       author = {{Dressler}, A. and {Gunn}, J.~E.},
        title = "{Spectroscopy of galaxies in distant clusters. II. The population of the 3C 295 cluster.}",
      journal = {\apj},
         year = 1983,
        month = jul,
       volume = {270},
        pages = {7-19},
          doi = {10.1086/161093},
       adsurl = {https://ui.adsabs.harvard.edu/abs/1983ApJ...270....7D}
}

@ARTICLE{Wild2020,
       author = {{Wild}, Vivienne and {Taj Aldeen}, Laith and {Carnall}, Adam and {Maltby}, David and {Almaini}, Omar and {Werle}, Ariel and {Wilkinson}, Aaron and {Rowlands}, Kate and {Bolzonella}, Micol and {Castellano}, Marco and {Gargiulo}, Adriana and {McLure}, Ross and {Pentericci}, Laura and {Pozzetti}, Lucia},
        title = "{The star formation histories of z {\ensuremath{\sim}} 1 post-starburst galaxies}",
      journal = {\mnras},
         year = 2020,
        month = may,
       volume = {494},
       number = {1},
        pages = {529-548},
          doi = {10.1093/mnras/staa674},
archivePrefix = {arXiv},
       eprint = {2001.09154},
 primaryClass = {astro-ph.GA},
       adsurl = {https://ui.adsabs.harvard.edu/abs/2020MNRAS.494..529W}
}

@ARTICLE{French2021,
       author = {{French}, K. Decker},
        title = "{Evolution Through the Post-starburst Phase: Using Post-starburst Galaxies as Laboratories for Understanding the Processes that Drive Galaxy Evolution.}",
      journal = {\pasp},
         year = 2021,
        month = jul,
       volume = {133},
       number = {1025},
          eid = {072001},
        pages = {072001},
          doi = {10.1088/1538-3873/ac0a59},
archivePrefix = {arXiv},
       eprint = {2106.05982},
 primaryClass = {astro-ph.GA},
       adsurl = {https://ui.adsabs.harvard.edu/abs/2021PASP..133g2001F}
}

@ARTICLE{Balogh1999,
       author = {{Balogh}, Michael L. and {Morris}, Simon L. and {Yee}, H.~K.~C. and {Carlberg}, R.~G. and {Ellingson}, Erica},
        title = "{Differential Galaxy Evolution in Cluster and Field Galaxies at z\raisebox{-0.5ex}\textasciitilde0.3}",
      journal = {\apj},
         year = 1999,
        month = dec,
       volume = {527},
       number = {1},
        pages = {54-79},
          doi = {10.1086/308056},
archivePrefix = {arXiv},
       eprint = {astro-ph/9906470},
 primaryClass = {astro-ph},
       adsurl = {https://ui.adsabs.harvard.edu/abs/1999ApJ...527...54B}
}

@ARTICLE{Tran2003,
       author = {{Tran}, Kim-Vy H. and {Franx}, Marijn and {Illingworth}, Garth and {Kelson}, Daniel D. and {van Dokkum}, Pieter},
        title = "{The Nature of E+A Galaxies in Intermediate-Redshift Clusters}",
      journal = {\apj},
         year = 2003,
        month = dec,
       volume = {599},
       number = {2},
        pages = {865-885},
          doi = {10.1086/379804},
archivePrefix = {arXiv},
       eprint = {astro-ph/0309460},
 primaryClass = {astro-ph},
       adsurl = {https://ui.adsabs.harvard.edu/abs/2003ApJ...599..865T}
}

@ARTICLE{Whitaker2012,
       author = {{Whitaker}, Katherine E. and {Kriek}, Mariska and {van Dokkum}, Pieter G. and {Bezanson}, Rachel and {Brammer}, Gabriel and {Franx}, Marijn and {Labb{\'e}}, Ivo},
        title = "{A Large Population of Massive Compact Post-starburst Galaxies at z > 1: Implications for the Size Evolution and Quenching Mechanism of Quiescent Galaxies}",
      journal = {\apj},
         year = 2012,
        month = feb,
       volume = {745},
       number = {2},
          eid = {179},
        pages = {179},
          doi = {10.1088/0004-637X/745/2/179},
archivePrefix = {arXiv},
       eprint = {1112.0313},
 primaryClass = {astro-ph.CO},
       adsurl = {https://ui.adsabs.harvard.edu/abs/2012ApJ...745..179W}
}

@ARTICLE{Akins2022,
       author = {{Akins}, Hollis B. and {Narayanan}, Desika and {Whitaker}, Katherine E. and {Dav{\'e}}, Romeel and {Lower}, Sidney and {Bezanson}, Rachel and {Feldmann}, Robert and {Kriek}, Mariska},
        title = "{Quenching and the UVJ Diagram in the SIMBA Cosmological Simulation}",
      journal = {\apj},
         year = 2022,
        month = apr,
       volume = {929},
       number = {1},
          eid = {94},
        pages = {94},
          doi = {10.3847/1538-4357/ac5d3a},
archivePrefix = {arXiv},
       eprint = {2105.12748},
 primaryClass = {astro-ph.GA},
       adsurl = {https://ui.adsabs.harvard.edu/abs/2022ApJ...929...94A}
}

@ARTICLE{Wild2014,
       author = {{Wild}, Vivienne and {Almaini}, Omar and {Cirasuolo}, Michele and {Dunlop}, Jim and {McLure}, Ross and {Bowler}, Rebecca and {Ferreira}, Joao and {Bradshaw}, Emma and {Chuter}, Robert and {Hartley}, Will},
        title = "{A new method for classifying galaxy SEDs from multiwavelength photometry}",
      journal = {\mnras},
         year = 2014,
        month = may,
       volume = {440},
       number = {2},
        pages = {1880-1898},
          doi = {10.1093/mnras/stu212},
archivePrefix = {arXiv},
       eprint = {1401.7878},
 primaryClass = {astro-ph.CO},
       adsurl = {https://ui.adsabs.harvard.edu/abs/2014MNRAS.440.1880W}
}

@ARTICLE{Maltby2016,
       author = {{Maltby}, David T. and {Almaini}, Omar and {Wild}, Vivienne and {Hatch}, Nina A. and {Hartley}, William G. and {Simpson}, Chris and {McLure}, Ross J. and {Dunlop}, James and {Rowlands}, Kate and {Cirasuolo}, Michele},
        title = "{The identification of post-starburst galaxies at z {\ensuremath{\sim}} 1 using multiwavelength photometry: a spectroscopic verification}",
      journal = {\mnras},
         year = 2016,
        month = jun,
       volume = {459},
       number = {1},
        pages = {L114-L118},
          doi = {10.1093/mnrasl/slw057},
archivePrefix = {arXiv},
       eprint = {1603.08941},
 primaryClass = {astro-ph.GA},
       adsurl = {https://ui.adsabs.harvard.edu/abs/2016MNRAS.459L.114M}
}

@ARTICLE{Wild2016,
       author = {{Wild}, Vivienne and {Almaini}, Omar and {Dunlop}, Jim and {Simpson}, Chris and {Rowlands}, Kate and {Bowler}, Rebecca and {Maltby}, David and {McLure}, Ross},
        title = "{The evolution of post-starburst galaxies from z=2 to 0.5}",
      journal = {\mnras},
         year = 2016,
        month = nov,
       volume = {463},
       number = {1},
        pages = {832-844},
          doi = {10.1093/mnras/stw1996},
archivePrefix = {arXiv},
       eprint = {1608.00588},
 primaryClass = {astro-ph.GA},
       adsurl = {https://ui.adsabs.harvard.edu/abs/2016MNRAS.463..832W}
}

@ARTICLE{Maltby2018,
       author = {{Maltby}, David T. and {Almaini}, Omar and {Wild}, Vivienne and {Hatch}, Nina A. and {Hartley}, William G. and {Simpson}, Chris and {Rowlands}, Kate and {Socolovsky}, Miguel},
        title = "{The structure of post-starburst galaxies at 0.5 < z < 2: evidence for two distinct quenching routes at different epochs}",
      journal = {\mnras},
         year = 2018,
        month = oct,
       volume = {480},
       number = {1},
        pages = {381-401},
          doi = {10.1093/mnras/sty1794},
archivePrefix = {arXiv},
       eprint = {1807.01325},
 primaryClass = {astro-ph.GA},
       adsurl = {https://ui.adsabs.harvard.edu/abs/2018MNRAS.480..381M}
}

@ARTICLE{Almaini2017,
       author = {{Almaini}, Omar and {Wild}, Vivienne and {Maltby}, David T. and {Hartley}, William G. and {Simpson}, Chris and {Hatch}, Nina A. and {McLure}, Ross J. and {Dunlop}, James S. and {Rowlands}, Kate},
        title = "{Massive post-starburst galaxies at z > 1 are compact proto-spheroids}",
      journal = {\mnras},
         year = 2017,
        month = dec,
       volume = {472},
       number = {2},
        pages = {1401-1412},
          doi = {10.1093/mnras/stx1957},
archivePrefix = {arXiv},
       eprint = {1708.00005},
 primaryClass = {astro-ph.GA},
       adsurl = {https://ui.adsabs.harvard.edu/abs/2017MNRAS.472.1401A}
}

@ARTICLE{Lanz2022,
       author = {{Lanz}, Lauranne and {Stepanoff}, Sofia and {Hickox}, Ryan C. and {Alatalo}, Katherine and {French}, K. Decker and {Rowlands}, Kate and {Nyland}, Kristina and {Appleton}, Philip N. and {Lacy}, Mark and {Medling}, Anne and {Mulchaey}, John S. and {Sazonova}, Elizaveta and {Urry}, Claudia Megan},
        title = "{Are Active Galactic Nuclei in Post-starburst Galaxies Driving the Change or Along for the Ride?}",
      journal = {\apj},
         year = 2022,
        month = aug,
       volume = {935},
       number = {1},
          eid = {29},
        pages = {29},
          doi = {10.3847/1538-4357/ac7d56},
archivePrefix = {arXiv},
       eprint = {2207.00607},
 primaryClass = {astro-ph.GA},
       adsurl = {https://ui.adsabs.harvard.edu/abs/2022ApJ...935...29L}
}

@ARTICLE{Tremonti2007,
       author = {{Tremonti}, Christy A. and {Moustakas}, John and {Diamond-Stanic}, Aleksandar M.},
        title = "{The Discovery of 1000 km s$^{-1}$ Outflows in Massive Poststarburst Galaxies at z=0.6}",
      journal = {\apjl},
         year = 2007,
        month = jul,
       volume = {663},
       number = {2},
        pages = {L77-L80},
          doi = {10.1086/520083},
archivePrefix = {arXiv},
       eprint = {0706.0527},
 primaryClass = {astro-ph},
       adsurl = {https://ui.adsabs.harvard.edu/abs/2007ApJ...663L..77T}
}

@ARTICLE{Maltby2019,
       author = {{Maltby}, David T. and {Almaini}, Omar and {McLure}, Ross J. and {Wild}, Vivienne and {Dunlop}, James and {Rowlands}, Kate and {Hartley}, William G. and {Hatch}, Nina A. and {Socolovsky}, Miguel and {Wilkinson}, Aaron and {Amorin}, Ricardo and {Bradshaw}, Emma J. and {Carnall}, Adam C. and {Castellano}, Marco and {Cimatti}, Andrea and {Cresci}, Giovanni and {Cullen}, Fergus and {De Barros}, Stephane and {Fontanot}, Fabio and {Garilli}, Bianca and {Koekemoer}, Anton M. and {McLeod}, Derek J. and {Pentericci}, Laura and {Talia}, Margherita},
        title = "{High-velocity outflows in massive post-starburst galaxies at z > 1}",
      journal = {\mnras},
         year = 2019,
        month = oct,
       volume = {489},
       number = {1},
        pages = {1139-1151},
          doi = {10.1093/mnras/stz2211},
archivePrefix = {arXiv},
       eprint = {1908.02766},
 primaryClass = {astro-ph.GA},
       adsurl = {https://ui.adsabs.harvard.edu/abs/2019MNRAS.489.1139M}
}

@ARTICLE{Taylor2024,
       author = {{Taylor}, Elizabeth and {Maltby}, David and {Almaini}, Omar and {Merrifield}, Michael and {Wild}, Vivienne and {Rowlands}, Kate and {Harrold}, Jimi},
        title = "{High-velocity outflows persist up to 1 Gyr after a starburst in recently quenched galaxies at z > 1}",
      journal = {\mnras},
         year = 2024,
        month = dec,
       volume = {535},
       number = {2},
        pages = {1684-1692},
          doi = {10.1093/mnras/stae2463},
archivePrefix = {arXiv},
       eprint = {2411.00102},
 primaryClass = {astro-ph.GA},
       adsurl = {https://ui.adsabs.harvard.edu/abs/2024MNRAS.535.1684T}
}

@ARTICLE{Polletta2007,
       author = {{Polletta}, M. and {Tajer}, M. and {Maraschi}, L. and {Trinchieri}, G. and {Lonsdale}, C.~J. and {Chiappetti}, L. and {Andreon}, S. and {Pierre}, M. and {Le F{\`e}vre}, O. and {Zamorani}, G. and {Maccagni}, D. and {Garcet}, O. and {Surdej}, J. and {Franceschini}, A. and {Alloin}, D. and {Shupe}, D.~L. and {Surace}, J.~A. and {Fang}, F. and {Rowan-Robinson}, M. and {Smith}, H.~E. and {Tresse}, L.},
        title = "{Spectral Energy Distributions of Hard X-Ray Selected Active Galactic Nuclei in the XMM-Newton Medium Deep Survey}",
      journal = {\apj},
         year = 2007,
        month = jul,
       volume = {663},
       number = {1},
        pages = {81-102},
          doi = {10.1086/518113},
archivePrefix = {arXiv},
       eprint = {astro-ph/0703255},
 primaryClass = {astro-ph},
       adsurl = {https://ui.adsabs.harvard.edu/abs/2007ApJ...663...81P}
}

@MISC{Dunlop2021,
       author = {{Dunlop}, James S. and {Abraham}, Roberto G. and {Ashby}, Matthew L.~N. and {Bagley}, Micaela and {Best}, Philip N. and {Bongiorno}, Angela and {Bouwens}, Rychard and {Bowler}, Rebecca A.~A. and {Brammer}, Gabriel and {Bremer}, Malcolm and {Calabro'}, Antonello and {Carnall}, Adam and {Castellano}, Marco and {Cirasuolo}, Michele and {Conselice}, Christopher and {Cullen}, Fergus and {Dave}, Romeel and {Dayal}, Pratika and {Dekel}, Avishai and {Dickinson}, Mark and {Duncan}, Kenneth James and {Elbaz}, David and {Ellis}, Richard S. and {Ferguson}, Harry C. and {Ferrara}, Andrea and {Finkelstein}, Steven L. and {Fontana}, Adriano and {Furlanetto}, Steven and {Fynbo}, Johan P.~U. and {Gallerani}, Simona and {Gardner}, Jonathan P. and {Giavalisco}, Mauro and {Grazian}, Andrea and {Grogin}, Norman and {Harikane}, Yuichi and {Hopkins}, Philip F. and {Ilbert}, Olivier and {Illingworth}, Garth D. and {Juneau}, Stephanie and {Jung}, Intae and {Kartaltepe}, Jeyhan and {Kassin}, Susan and {Kauffmann}, Olivier Benjamin and {Khochfar}, Sadegh and {Kirkpatrick}, Allison and {Kocevski}, Dale D. and {Koekemoer}, Anton M. and {Labbe}, Ivo and {Laporte}, Nicolas and {Larson}, Rebecca L. and {Lucas}, Ray A. and {Magee}, Daniel K. and {Mason}, Charlotte and {McCracken}, Henry Joy and {McLeod}, Derek and {McLure}, Ross and {Merlin}, Emiliano and {Mesinger}, Andrei and {Milvang-Jensen}, Bo and {Newman}, Jeffrey Allen and {Oesch}, Pascal and {Ouchi}, Masami and {Pacifici}, Camilla and {Papovich}, Casey and {Peacock}, John and {Peeples}, Molly and {Pentericci}, Laura and {Perez-Gonzalez}, Pablo G. and {Pirzkal}, Norbert and {Pope}, Alexandra and {Pye}, John P. and {Reddy}, Naveen A. and {Robertson}, Brant and {Salvato}, Mara and {Santini}, Paola and {Schaerer}, Daniel and {Shapley}, Alice E. and {Simons}, Raymond and {Smit}, Renske and {Smith}, Britton D. and {Snyder}, Greg and {Somerville}, Rachel S. and {Stanway}, Elizabeth R. and {Stefanon}, Mauro and {Tasca}, Lidia and {Tikkanen}, Tuomo and {Tresse}, Laurence and {Trump}, Jonathan R. and {Whitaker}, Katherine E. and {Wilkins}, Stephen Matthew and {Wright}, Gillian and {Wyithe}, J. Stuart B. and {van Dokkum}, Pieter and {van der Werf}, Paul},
        title = "{PRIMER: Public Release IMaging for Extragalactic Research}",
 howpublished = {JWST Proposal. Cycle 1, ID. \#1837},
         year = 2021,
        month = mar,
        pages = {1837},
       adsurl = {https://ui.adsabs.harvard.edu/abs/2021jwst.prop.1837D}
}

@ARTICLE{Donnan2024,
       author = {{Donnan}, C.~T. and {McLure}, R.~J. and {Dunlop}, J.~S. and {McLeod}, D.~J. and {Magee}, D. and {Arellano-C{\'o}rdova}, K.~Z. and {Barrufet}, L. and {Begley}, R. and {Bowler}, R.~A.~A. and {Carnall}, A.~C. and {Cullen}, F. and {Ellis}, R.~S. and {Fontana}, A. and {Illingworth}, G.~D. and {Grogin}, N.~A. and {Hamadouche}, M.~L. and {Koekemoer}, A.~M. and {Liu}, F. -Y. and {Mason}, C. and {Santini}, P. and {Stanton}, T.~M.},
        title = "{JWST PRIMER: a new multifield determination of the evolving galaxy UV luminosity function at redshifts z ≃ 9 - 15}",
      journal = {\mnras},
         year = 2024,
        month = sep,
       volume = {533},
       number = {3},
        pages = {3222-3237},
          doi = {10.1093/mnras/stae2037},
archivePrefix = {arXiv},
       eprint = {2403.03171},
 primaryClass = {astro-ph.GA},
       adsurl = {https://ui.adsabs.harvard.edu/abs/2024MNRAS.533.3222D}
}

@ARTICLE{Grogin2011,
       author = {{Grogin}, Norman A. and {Kocevski}, Dale D. and {Faber}, S.~M. and {Ferguson}, Henry C. and {Koekemoer}, Anton M. and {Riess}, Adam G. and {Acquaviva}, Viviana and {Alexander}, David M. and {Almaini}, Omar and {Ashby}, Matthew L.~N. and {Barden}, Marco and {Bell}, Eric F. and {Bournaud}, Fr{\'e}d{\'e}ric and {Brown}, Thomas M. and {Caputi}, Karina I. and {Casertano}, Stefano and {Cassata}, Paolo and {Castellano}, Marco and {Challis}, Peter and {Chary}, Ranga-Ram and {Cheung}, Edmond and {Cirasuolo}, Michele and {Conselice}, Christopher J. and {Roshan Cooray}, Asantha and {Croton}, Darren J. and {Daddi}, Emanuele and {Dahlen}, Tomas and {Dav{\'e}}, Romeel and {de Mello}, Du{\'\i}lia F. and {Dekel}, Avishai and {Dickinson}, Mark and {Dolch}, Timothy and {Donley}, Jennifer L. and {Dunlop}, James S. and {Dutton}, Aaron A. and {Elbaz}, David and {Fazio}, Giovanni G. and {Filippenko}, Alexei V. and {Finkelstein}, Steven L. and {Fontana}, Adriano and {Gardner}, Jonathan P. and {Garnavich}, Peter M. and {Gawiser}, Eric and {Giavalisco}, Mauro and {Grazian}, Andrea and {Guo}, Yicheng and {Hathi}, Nimish P. and {H{\"a}ussler}, Boris and {Hopkins}, Philip F. and {Huang}, Jia-Sheng and {Huang}, Kuang-Han and {Jha}, Saurabh W. and {Kartaltepe}, Jeyhan S. and {Kirshner}, Robert P. and {Koo}, David C. and {Lai}, Kamson and {Lee}, Kyoung-Soo and {Li}, Weidong and {Lotz}, Jennifer M. and {Lucas}, Ray A. and {Madau}, Piero and {McCarthy}, Patrick J. and {McGrath}, Elizabeth J. and {McIntosh}, Daniel H. and {McLure}, Ross J. and {Mobasher}, Bahram and {Moustakas}, Leonidas A. and {Mozena}, Mark and {Nandra}, Kirpal and {Newman}, Jeffrey A. and {Niemi}, Sami-Matias and {Noeske}, Kai G. and {Papovich}, Casey J. and {Pentericci}, Laura and {Pope}, Alexandra and {Primack}, Joel R. and {Rajan}, Abhijith and {Ravindranath}, Swara and {Reddy}, Naveen A. and {Renzini}, Alvio and {Rix}, Hans-Walter and {Robaina}, Aday R. and {Rodney}, Steven A. and {Rosario}, David J. and {Rosati}, Piero and {Salimbeni}, Sara and {Scarlata}, Claudia and {Siana}, Brian and {Simard}, Luc and {Smidt}, Joseph and {Somerville}, Rachel S. and {Spinrad}, Hyron and {Straughn}, Amber N. and {Strolger}, Louis-Gregory and {Telford}, Olivia and {Teplitz}, Harry I. and {Trump}, Jonathan R. and {van der Wel}, Arjen and {Villforth}, Carolin and {Wechsler}, Risa H. and {Weiner}, Benjamin J. and {Wiklind}, Tommy and {Wild}, Vivienne and {Wilson}, Grant and {Wuyts}, Stijn and {Yan}, Hao-Jing and {Yun}, Min S.},
        title = "{CANDELS: The Cosmic Assembly Near-infrared Deep Extragalactic Legacy Survey}",
      journal = {\apjs},
         year = 2011,
        month = dec,
       volume = {197},
       number = {2},
          eid = {35},
        pages = {35},
          doi = {10.1088/0067-0049/197/2/35},
archivePrefix = {arXiv},
       eprint = {1105.3753},
 primaryClass = {astro-ph.CO},
       adsurl = {https://ui.adsabs.harvard.edu/abs/2011ApJS..197...35G}
}

@ARTICLE{Pozzetti2010,
       author = {{Pozzetti}, L. and {Bolzonella}, M. and {Zucca}, E. and {Zamorani}, G. and {Lilly}, S. and {Renzini}, A. and {Moresco}, M. and {Mignoli}, M. and {Cassata}, P. and {Tasca}, L. and {Lamareille}, F. and {Maier}, C. and {Meneux}, B. and {Halliday}, C. and {Oesch}, P. and {Vergani}, D. and {Caputi}, K. and {Kova{\v{c}}}, K. and {Cimatti}, A. and {Cucciati}, O. and {Iovino}, A. and {Peng}, Y. and {Carollo}, M. and {Contini}, T. and {Kneib}, J. -P. and {Le F{\'e}vre}, O. and {Mainieri}, V. and {Scodeggio}, M. and {Bardelli}, S. and {Bongiorno}, A. and {Coppa}, G. and {de la Torre}, S. and {de Ravel}, L. and {Franzetti}, P. and {Garilli}, B. and {Kampczyk}, P. and {Knobel}, C. and {Le Borgne}, J. -F. and {Le Brun}, V. and {Pell{\`o}}, R. and {Perez Montero}, E. and {Ricciardelli}, E. and {Silverman}, J.~D. and {Tanaka}, M. and {Tresse}, L. and {Abbas}, U. and {Bottini}, D. and {Cappi}, A. and {Guzzo}, L. and {Koekemoer}, A.~M. and {Leauthaud}, A. and {Maccagni}, D. and {Marinoni}, C. and {McCracken}, H.~J. and {Memeo}, P. and {Porciani}, C. and {Scaramella}, R. and {Scarlata}, C. and {Scoville}, N.},
        title = "{zCOSMOS - 10k-bright spectroscopic sample. The bimodality in the galaxy stellar mass function: exploring its evolution with redshift}",
      journal = {\aap},
         year = 2010,
        month = nov,
       volume = {523},
          eid = {A13},
        pages = {A13},
          doi = {10.1051/0004-6361/200913020},
archivePrefix = {arXiv},
       eprint = {0907.5416},
 primaryClass = {astro-ph.CO},
       adsurl = {https://ui.adsabs.harvard.edu/abs/2010A&A...523A..13P}
}

@ARTICLE{sextractor,
       author = {{Bertin}, E. and {Arnouts}, S.},
        title = "{SExtractor: Software for source extraction.}",
      journal = {\aaps},
         year = 1996,
        month = jun,
       volume = {117},
        pages = {393-404},
          doi = {10.1051/aas:1996164},
       adsurl = {https://ui.adsabs.harvard.edu/abs/1996A&AS..117..393B}
}

@ARTICLE{EAZY,
       author = {{Brammer}, Gabriel B. and {van Dokkum}, Pieter G. and {Coppi}, Paolo},
        title = "{EAZY: A Fast, Public Photometric Redshift Code}",
      journal = {\apj},
         year = 2008,
        month = oct,
       volume = {686},
       number = {2},
        pages = {1503-1513},
          doi = {10.1086/591786},
archivePrefix = {arXiv},
       eprint = {0807.1533},
 primaryClass = {astro-ph},
       adsurl = {https://ui.adsabs.harvard.edu/abs/2008ApJ...686.1503B}
}

@ARTICLE{Bradshaw2013,
       author = {{Bradshaw}, E.~J. and {Almaini}, O. and {Hartley}, W.~G. and {Smith}, K.~T. and {Conselice}, C.~J. and {Dunlop}, J.~S. and {Simpson}, C. and {Chuter}, R.~W. and {Cirasuolo}, M. and {Foucaud}, S. and {McLure}, R.~J. and {Mortlock}, A. and {Pearce}, H.},
        title = "{High-velocity outflows from young star-forming galaxies in the UKIDSS Ultra-Deep Survey}",
      journal = {\mnras},
         year = 2013,
        month = jul,
       volume = {433},
       number = {1},
        pages = {194-208},
          doi = {10.1093/mnras/stt715},
archivePrefix = {arXiv},
       eprint = {1304.7276},
 primaryClass = {astro-ph.CO},
       adsurl = {https://ui.adsabs.harvard.edu/abs/2013MNRAS.433..194B}
}

@ARTICLE{McLure2013,
       author = {{McLure}, R.~J. and {Pearce}, H.~J. and {Dunlop}, J.~S. and {Cirasuolo}, M. and {Curtis-Lake}, E. and {Bruce}, V.~A. and {Caputi}, K.~I. and {Almaini}, O. and {Bonfield}, D.~G. and {Bradshaw}, E.~J. and {Buitrago}, F. and {Chuter}, R. and {Foucaud}, S. and {Hartley}, W.~G. and {Jarvis}, M.~J.},
        title = "{The sizes, masses and specific star formation rates of massive galaxies at 1.3 < z < 1.5: strong evidence in favour of evolution via minor mergers}",
      journal = {\mnras},
         year = 2013,
        month = jan,
       volume = {428},
       number = {2},
        pages = {1088-1106},
          doi = {10.1093/mnras/sts092},
archivePrefix = {arXiv},
       eprint = {1205.4058},
 primaryClass = {astro-ph.CO},
       adsurl = {https://ui.adsabs.harvard.edu/abs/2013MNRAS.428.1088M}
}

@ARTICLE{McLure2018,
       author = {{McLure}, R.~J. and {Pentericci}, L. and {Cimatti}, A. and {Dunlop}, J.~S. and {Elbaz}, D. and {Fontana}, A. and {Nandra}, K. and {Amorin}, R. and {Bolzonella}, M. and {Bongiorno}, A. and {Carnall}, A.~C. and {Castellano}, M. and {Cirasuolo}, M. and {Cucciati}, O. and {Cullen}, F. and {De Barros}, S. and {Finkelstein}, S.~L. and {Fontanot}, F. and {Franzetti}, P. and {Fumana}, M. and {Gargiulo}, A. and {Garilli}, B. and {Guaita}, L. and {Hartley}, W.~G. and {Iovino}, A. and {Jarvis}, M.~J. and {Juneau}, S. and {Karman}, W. and {Maccagni}, D. and {Marchi}, F. and {M{\'a}rmol-Queralt{\'o}}, E. and {Pompei}, E. and {Pozzetti}, L. and {Scodeggio}, M. and {Sommariva}, V. and {Talia}, M. and {Almaini}, O. and {Balestra}, I. and {Bardelli}, S. and {Bell}, E.~F. and {Bourne}, N. and {Bowler}, R.~A.~A. and {Brusa}, M. and {Buitrago}, F. and {Caputi}, K.~I. and {Cassata}, P. and {Charlot}, S. and {Citro}, A. and {Cresci}, G. and {Cristiani}, S. and {Curtis-Lake}, E. and {Dickinson}, M. and {Fazio}, G.~G. and {Ferguson}, H.~C. and {Fiore}, F. and {Franco}, M. and {Fynbo}, J.~P.~U. and {Galametz}, A. and {Georgakakis}, A. and {Giavalisco}, M. and {Grazian}, A. and {Hathi}, N.~P. and {Jung}, I. and {Kim}, S. and {Koekemoer}, A.~M. and {Khusanova}, Y. and {Le F{\`e}vre}, O. and {Lotz}, J.~M. and {Mannucci}, F. and {Maltby}, D.~T. and {Matsuoka}, K. and {McLeod}, D.~J. and {Mendez-Hernandez}, H. and {Mendez-Abreu}, J. and {Mignoli}, M. and {Moresco}, M. and {Mortlock}, A. and {Nonino}, M. and {Pannella}, M. and {Papovich}, C. and {Popesso}, P. and {Rosario}, D.~P. and {Salvato}, M. and {Santini}, P. and {Schaerer}, D. and {Schreiber}, C. and {Stark}, D.~P. and {Tasca}, L.~A.~M. and {Thomas}, R. and {Treu}, T. and {Vanzella}, E. and {Wild}, V. and {Williams}, C.~C. and {Zamorani}, G. and {Zucca}, E.},
        title = "{The VANDELS ESO public spectroscopic survey}",
      journal = {\mnras},
         year = 2018,
        month = sep,
       volume = {479},
       number = {1},
        pages = {25-42},
          doi = {10.1093/mnras/sty1213},
archivePrefix = {arXiv},
       eprint = {1803.07414},
 primaryClass = {astro-ph.GA},
       adsurl = {https://ui.adsabs.harvard.edu/abs/2018MNRAS.479...25M}
}

@ARTICLE{Pentericci2018,
       author = {{Pentericci}, L. and {McLure}, R.~J. and {Garilli}, B. and {Cucciati}, O. and {Franzetti}, P. and {Iovino}, A. and {Amorin}, R. and {Bolzonella}, M. and {Bongiorno}, A. and {Carnall}, A.~C. and {Castellano}, M. and {Cimatti}, A. and {Cirasuolo}, M. and {Cullen}, F. and {De Barros}, S. and {Dunlop}, J.~S. and {Elbaz}, D. and {Finkelstein}, S.~L. and {Fontana}, A. and {Fontanot}, F. and {Fumana}, M. and {Gargiulo}, A. and {Guaita}, L. and {Hartley}, W.~G. and {Jarvis}, M.~J. and {Juneau}, S. and {Karman}, W. and {Maccagni}, D. and {Marchi}, F. and {Marmol-Queralto}, E. and {Nandra}, K. and {Pompei}, E. and {Pozzetti}, L. and {Scodeggio}, M. and {Sommariva}, V. and {Talia}, M. and {Almaini}, O. and {Balestra}, I. and {Bardelli}, S. and {Bell}, E.~F. and {Bourne}, N. and {Bowler}, R.~A.~A. and {Brusa}, M. and {Buitrago}, F. and {Caputi}, K.~I. and {Cassata}, P. and {Charlot}, S. and {Citro}, A. and {Cresci}, G. and {Cristiani}, S. and {Curtis-Lake}, E. and {Dickinson}, M. and {Fazio}, G.~G. and {Ferguson}, H.~C. and {Fiore}, F. and {Franco}, M. and {Fynbo}, J.~P.~U. and {Galametz}, A. and {Georgakakis}, A. and {Giavalisco}, M. and {Grazian}, A. and {Hathi}, N.~P. and {Jung}, I. and {Kim}, S. and {Koekemoer}, A.~M. and {Khusanova}, Y. and {Le F{\`e}vre}, O. and {Lotz}, J.~M. and {Mannucci}, F. and {Maltby}, D.~T. and {Matsuoka}, K. and {McLeod}, D.~J. and {Mendez-Hernandez}, H. and {Mendez-Abreu}, J. and {Mignoli}, M. and {Moresco}, M. and {Mortlock}, A. and {Nonino}, M. and {Pannella}, M. and {Papovich}, C. and {Popesso}, P. and {Rosario}, D.~P. and {Salvato}, M. and {Santini}, P. and {Schaerer}, D. and {Schreiber}, C. and {Stark}, D.~P. and {Tasca}, L.~A.~M. and {Thomas}, R. and {Treu}, T. and {Vanzella}, E. and {Wild}, V. and {Williams}, C.~C. and {Zamorani}, G. and {Zucca}, E.},
        title = "{The VANDELS ESO public spectroscopic survey: Observations and first data release}",
      journal = {\aap},
         year = 2018,
        month = sep,
       volume = {616},
          eid = {A174},
        pages = {A174},
          doi = {10.1051/0004-6361/201833047},
archivePrefix = {arXiv},
       eprint = {1803.07373},
 primaryClass = {astro-ph.GA},
       adsurl = {https://ui.adsabs.harvard.edu/abs/2018A&A...616A.174P}
}

@ARTICLE{Carnall2024,
       author = {{Carnall}, A.~C. and {Cullen}, F. and {McLure}, R.~J. and {McLeod}, D.~J. and {Begley}, R. and {Donnan}, C.~T. and {Dunlop}, J.~S. and {Shapley}, A.~E. and {Rowlands}, K. and {Almaini}, O. and {Arellano-C{\'o}rdova}, K.~Z. and {Barrufet}, L. and {Cimatti}, A. and {Ellis}, R.~S. and {Grogin}, N.~A. and {Hamadouche}, M.~L. and {Illingworth}, G.~D. and {Koekemoer}, A.~M. and {Leung}, H. -H. and {Lovell}, C.~C. and {P{\'e}rez-Gonz{\'a}lez}, P.~G. and {Santini}, P. and {Stanton}, T.~M. and {Wild}, V.},
        title = "{The JWST EXCELS survey: too much, too young, too fast? Ultra-massive quiescent galaxies at 3 < z < 5}",
      journal = {\mnras},
         year = 2024,
        month = oct,
       volume = {534},
       number = {1},
        pages = {325-348},
          doi = {10.1093/mnras/stae2092},
archivePrefix = {arXiv},
       eprint = {2405.02242},
 primaryClass = {astro-ph.GA},
       adsurl = {https://ui.adsabs.harvard.edu/abs/2024MNRAS.534..325C}
}

@ARTICLE{Wilman2008,
       author = {{Wilman}, D.~J. and {Pierini}, D. and {Tyler}, K. and {McGee}, S.~L. and {Oemler}, Jr., A. and {Morris}, S.~L. and {Balogh}, M.~L. and {Bower}, R.~G. and {Mulchaey}, J.~S.},
        title = "{Unveiling the Important Role of Groups in the Evolution of Massive Galaxies: Insights from an Infrared Passive Sequence at Intermediate Redshift}",
      journal = {\apj},
         year = 2008,
        month = jun,
       volume = {680},
       number = {2},
        pages = {1009-1021},
          doi = {10.1086/587478},
archivePrefix = {arXiv},
       eprint = {0802.2549},
 primaryClass = {astro-ph},
       adsurl = {https://ui.adsabs.harvard.edu/abs/2008ApJ...680.1009W}
}

@ARTICLE{Li2007,
       author = {{Li}, Hai-Ning and {Wu}, Hong and {Cao}, Chen and {Zhu}, Yi-Nan},
        title = "{Morphological Dependence of Mid-Infrared Properties of SDSS Galaxies in the Spitzer SWIRE Survey}",
      journal = {\aj},
         year = 2007,
        month = oct,
       volume = {134},
       number = {4},
        pages = {1315-1329},
          doi = {10.1086/520807},
archivePrefix = {arXiv},
       eprint = {0706.2712},
 primaryClass = {astro-ph},
       adsurl = {https://ui.adsabs.harvard.edu/abs/2007AJ....134.1315L}
}

@ARTICLE{Yang2022,
       author = {{Yang}, Guang and {Boquien}, M{\'e}d{\'e}ric and {Brandt}, W.~N. and {Buat}, V{\'e}ronique and {Burgarella}, Denis and {Ciesla}, Laure and {Lehmer}, Bret D. and {Ma{\l}ek}, Katarzyna and {Mountrichas}, George and {Papovich}, Casey and {Pons}, Estelle and {Stalevski}, Marko and {Theul{\'e}}, Patrice and {Zhu}, Shifu},
        title = "{Fitting AGN/Galaxy X-Ray-to-radio SEDs with CIGALE and Improvement of the Code}",
      journal = {\apj},
         year = 2022,
        month = mar,
       volume = {927},
       number = {2},
          eid = {192},
        pages = {192},
          doi = {10.3847/1538-4357/ac4971},
archivePrefix = {arXiv},
       eprint = {2201.03718},
 primaryClass = {astro-ph.GA},
       adsurl = {https://ui.adsabs.harvard.edu/abs/2022ApJ...927..192Y}
}

@ARTICLE{Stalevski2012,
       author = {{Stalevski}, Marko},
        title = "{SKIRTOR - database of modelled AGN dusty torus SEDs}",
      journal = {Bulgarian Astronomical Journal},
         year = 2012,
        month = sep,
       volume = {18},
       number = {3},
        pages = {3},
       adsurl = {https://ui.adsabs.harvard.edu/abs/2012BlgAJ..18c...3S}
}

@ARTICLE{Stalevski2016,
       author = {{Stalevski}, Marko and {Ricci}, Claudio and {Ueda}, Yoshihiro and {Lira}, Paulina and {Fritz}, Jacopo and {Baes}, Maarten},
        title = "{The dust covering factor in active galactic nuclei}",
      journal = {\mnras},
         year = 2016,
        month = may,
       volume = {458},
       number = {3},
        pages = {2288-2302},
          doi = {10.1093/mnras/stw444},
archivePrefix = {arXiv},
       eprint = {1602.06954},
 primaryClass = {astro-ph.GA},
       adsurl = {https://ui.adsabs.harvard.edu/abs/2016MNRAS.458.2288S}
}

@ARTICLE{BruzalCharlot2003,
       author = {{Bruzual}, G. and {Charlot}, S.},
        title = "{Stellar population synthesis at the resolution of 2003}",
      journal = {\mnras},
         year = 2003,
        month = oct,
       volume = {344},
       number = {4},
        pages = {1000-1028},
          doi = {10.1046/j.1365-8711.2003.06897.x},
archivePrefix = {arXiv},
       eprint = {astro-ph/0309134},
 primaryClass = {astro-ph},
       adsurl = {https://ui.adsabs.harvard.edu/abs/2003MNRAS.344.1000B}
}

@ARTICLE{Chabrier2003,
       author = {{Chabrier}, Gilles},
        title = "{Galactic Stellar and Substellar Initial Mass Function}",
      journal = {\pasp},
         year = 2003,
        month = jul,
       volume = {115},
       number = {809},
        pages = {763-795},
          doi = {10.1086/376392},
archivePrefix = {arXiv},
       eprint = {astro-ph/0304382},
 primaryClass = {astro-ph},
       adsurl = {https://ui.adsabs.harvard.edu/abs/2003PASP..115..763C}
}

@ARTICLE{Rieke2025,
       author = {{Rieke}, George H. and {Sun}, Yang and {Lyu}, Jianwei and {Willmer}, Christopher N.~A. and {Zhu}, Yongda and {Rinaldi}, Pierluigi and {Stone}, Meredith A. and {Hainline}, Kevin N. and {P{\'e}rez-Gonz{\'a}lez}, Pablo G.},
        title = "{Confirming Near- to Mid-infrared Photometrically Identified Obscured AGNs in the JWST Era}",
      journal = {\apj},
         year = 2025,
        month = nov,
       volume = {994},
       number = {1},
          eid = {35},
        pages = {35},
          doi = {10.3847/1538-4357/adff79},
archivePrefix = {arXiv},
       eprint = {2510.07303},
 primaryClass = {astro-ph.GA},
       adsurl = {https://ui.adsabs.harvard.edu/abs/2025ApJ...994...35R}
}

@ARTICLE{Kirkpatrick2012,
       author = {{Kirkpatrick}, Allison and {Pope}, Alexandra and {Alexander}, David M. and {Charmandaris}, Vassilis and {Daddi}, Emmanuele and {Dickinson}, Mark and {Elbaz}, David and {Gabor}, Jared and {Hwang}, Ho Seong and {Ivison}, Rob and {Mullaney}, James and {Pannella}, Maurilio and {Scott}, Douglas and {Altieri}, Bruno and {Aussel}, Herve and {Bournaud}, Fr{\'e}d{\'e}ric and {Buat}, Veronique and {Coia}, Daniela and {Dannerbauer}, Helmut and {Dasyra}, Kalliopi and {Kartaltepe}, Jeyhan and {Leiton}, Roger and {Lin}, Lihwai and {Magdis}, Georgios and {Magnelli}, Benjamin and {Morrison}, Glenn and {Popesso}, Paola and {Valtchanov}, Ivan},
        title = "{GOODS-Herschel: Impact of Active Galactic Nuclei and Star Formation Activity on Infrared Spectral Energy Distributions at High Redshift}",
      journal = {\apj},
         year = 2012,
        month = nov,
       volume = {759},
       number = {2},
          eid = {139},
        pages = {139},
          doi = {10.1088/0004-637X/759/2/139},
archivePrefix = {arXiv},
       eprint = {1209.4902},
 primaryClass = {astro-ph.CO},
       adsurl = {https://ui.adsabs.harvard.edu/abs/2012ApJ...759..139K}
}

@ARTICLE{Peeters2002,
       author = {{Peeters}, E. and {Hony}, S. and {Van Kerckhoven}, C. and {Tielens}, A.~G.~G.~M. and {Allamandola}, L.~J. and {Hudgins}, D.~M. and {Bauschlicher}, C.~W.},
        title = "{The rich 6 to 9 vec mu m spectrum of interstellar PAHs}",
      journal = {\aap},
         year = 2002,
        month = aug,
       volume = {390},
        pages = {1089-1113},
          doi = {10.1051/0004-6361:20020773},
archivePrefix = {arXiv},
       eprint = {astro-ph/0205400},
 primaryClass = {astro-ph},
       adsurl = {https://ui.adsabs.harvard.edu/abs/2002A&A...390.1089P}
}

@ARTICLE{Tielens2008,
       author = {{Tielens}, A.~G.~G.~M.},
        title = "{Interstellar polycyclic aromatic hydrocarbon molecules.}",
      journal = {\araa},
         year = 2008,
        month = sep,
       volume = {46},
        pages = {289-337},
          doi = {10.1146/annurev.astro.46.060407.145211},
       adsurl = {https://ui.adsabs.harvard.edu/abs/2008ARA&A..46..289T}
}

@ARTICLE{Smith2007,
       author = {{Smith}, J.~D.~T. and {Draine}, B.~T. and {Dale}, D.~A. and {Moustakas}, J. and {Kennicutt}, Jr., R.~C. and {Helou}, G. and {Armus}, L. and {Roussel}, H. and {Sheth}, K. and {Bendo}, G.~J. and {Buckalew}, B.~A. and {Calzetti}, D. and {Engelbracht}, C.~W. and {Gordon}, K.~D. and {Hollenbach}, D.~J. and {Li}, A. and {Malhotra}, S. and {Murphy}, E.~J. and {Walter}, F.},
        title = "{The Mid-Infrared Spectrum of Star-forming Galaxies: Global Properties of Polycyclic Aromatic Hydrocarbon Emission}",
      journal = {\apj},
         year = 2007,
        month = feb,
       volume = {656},
       number = {2},
        pages = {770-791},
          doi = {10.1086/510549},
archivePrefix = {arXiv},
       eprint = {astro-ph/0610913},
 primaryClass = {astro-ph},
       adsurl = {https://ui.adsabs.harvard.edu/abs/2007ApJ...656..770S}
}

@ARTICLE{Calzetti2007,
       author = {{Calzetti}, D. and {Kennicutt}, R.~C. and {Engelbracht}, C.~W. and {Leitherer}, C. and {Draine}, B.~T. and {Kewley}, L. and {Moustakas}, J. and {Sosey}, M. and {Dale}, D.~A. and {Gordon}, K.~D. and {Helou}, G.~X. and {Hollenbach}, D.~J. and {Armus}, L. and {Bendo}, G. and {Bot}, C. and {Buckalew}, B. and {Jarrett}, T. and {Li}, A. and {Meyer}, M. and {Murphy}, E.~J. and {Prescott}, M. and {Regan}, M.~W. and {Rieke}, G.~H. and {Roussel}, H. and {Sheth}, K. and {Smith}, J.~D.~T. and {Thornley}, M.~D. and {Walter}, F.},
        title = "{The Calibration of Mid-Infrared Star Formation Rate Indicators}",
      journal = {\apj},
         year = 2007,
        month = sep,
       volume = {666},
       number = {2},
        pages = {870-895},
          doi = {10.1086/520082},
archivePrefix = {arXiv},
       eprint = {0705.3377},
 primaryClass = {astro-ph},
       adsurl = {https://ui.adsabs.harvard.edu/abs/2007ApJ...666..870C}
}

@ARTICLE{Fabian1999,
       author = {{Fabian}, A.~C.},
        title = "{The obscured growth of massive black holes}",
      journal = {\mnras},
         year = 1999,
        month = oct,
       volume = {308},
       number = {4},
        pages = {L39-L43},
          doi = {10.1046/j.1365-8711.1999.03017.x},
archivePrefix = {arXiv},
       eprint = {astro-ph/9908064},
 primaryClass = {astro-ph},
       adsurl = {https://ui.adsabs.harvard.edu/abs/1999MNRAS.308L..39F}
}

@ARTICLE{Wild2010,
       author = {{Wild}, Vivienne and {Heckman}, Timothy and {Charlot}, St{\'e}phane},
        title = "{Timing the starburst-AGN connection}",
      journal = {\mnras},
         year = 2010,
        month = jun,
       volume = {405},
       number = {2},
        pages = {933-947},
          doi = {10.1111/j.1365-2966.2010.16536.x},
archivePrefix = {arXiv},
       eprint = {1002.3156},
 primaryClass = {astro-ph.CO},
       adsurl = {https://ui.adsabs.harvard.edu/abs/2010MNRAS.405..933W}
}

@ARTICLE{Koekemoer2011,
       author = {{Koekemoer}, Anton M. and {Faber}, S.~M. and {Ferguson}, Henry C. and {Grogin}, Norman A. and {Kocevski}, Dale D. and {Koo}, David C. and {Lai}, Kamson and {Lotz}, Jennifer M. and {Lucas}, Ray A. and {McGrath}, Elizabeth J. and {Ogaz}, Sara and {Rajan}, Abhijith and {Riess}, Adam G. and {Rodney}, Steve A. and {Strolger}, Louis and {Casertano}, Stefano and {Castellano}, Marco and {Dahlen}, Tomas and {Dickinson}, Mark and {Dolch}, Timothy and {Fontana}, Adriano and {Giavalisco}, Mauro and {Grazian}, Andrea and {Guo}, Yicheng and {Hathi}, Nimish P. and {Huang}, Kuang-Han and {van der Wel}, Arjen and {Yan}, Hao-Jing and {Acquaviva}, Viviana and {Alexander}, David M. and {Almaini}, Omar and {Ashby}, Matthew L.~N. and {Barden}, Marco and {Bell}, Eric F. and {Bournaud}, Fr{\'e}d{\'e}ric and {Brown}, Thomas M. and {Caputi}, Karina I. and {Cassata}, Paolo and {Challis}, Peter J. and {Chary}, Ranga-Ram and {Cheung}, Edmond and {Cirasuolo}, Michele and {Conselice}, Christopher J. and {Roshan Cooray}, Asantha and {Croton}, Darren J. and {Daddi}, Emanuele and {Dav{\'e}}, Romeel and {de Mello}, Duilia F. and {de Ravel}, Loic and {Dekel}, Avishai and {Donley}, Jennifer L. and {Dunlop}, James S. and {Dutton}, Aaron A. and {Elbaz}, David and {Fazio}, Giovanni G. and {Filippenko}, Alexei V. and {Finkelstein}, Steven L. and {Frazer}, Chris and {Gardner}, Jonathan P. and {Garnavich}, Peter M. and {Gawiser}, Eric and {Gruetzbauch}, Ruth and {Hartley}, Will G. and {H{\"a}ussler}, Boris and {Herrington}, Jessica and {Hopkins}, Philip F. and {Huang}, Jia-Sheng and {Jha}, Saurabh W. and {Johnson}, Andrew and {Kartaltepe}, Jeyhan S. and {Khostovan}, Ali A. and {Kirshner}, Robert P. and {Lani}, Caterina and {Lee}, Kyoung-Soo and {Li}, Weidong and {Madau}, Piero and {McCarthy}, Patrick J. and {McIntosh}, Daniel H. and {McLure}, Ross J. and {McPartland}, Conor and {Mobasher}, Bahram and {Moreira}, Heidi and {Mortlock}, Alice and {Moustakas}, Leonidas A. and {Mozena}, Mark and {Nandra}, Kirpal and {Newman}, Jeffrey A. and {Nielsen}, Jennifer L. and {Niemi}, Sami and {Noeske}, Kai G. and {Papovich}, Casey J. and {Pentericci}, Laura and {Pope}, Alexandra and {Primack}, Joel R. and {Ravindranath}, Swara and {Reddy}, Naveen A. and {Renzini}, Alvio and {Rix}, Hans-Walter and {Robaina}, Aday R. and {Rosario}, David J. and {Rosati}, Piero and {Salimbeni}, Sara and {Scarlata}, Claudia and {Siana}, Brian and {Simard}, Luc and {Smidt}, Joseph and {Snyder}, Diana and {Somerville}, Rachel S. and {Spinrad}, Hyron and {Straughn}, Amber N. and {Telford}, Olivia and {Teplitz}, Harry I. and {Trump}, Jonathan R. and {Vargas}, Carlos and {Villforth}, Carolin and {Wagner}, Cory R. and {Wandro}, Pat and {Wechsler}, Risa H. and {Weiner}, Benjamin J. and {Wiklind}, Tommy and {Wild}, Vivienne and {Wilson}, Grant and {Wuyts}, Stijn and {Yun}, Min S.},
        title = "{CANDELS: The Cosmic Assembly Near-infrared Deep Extragalactic Legacy Survey{\textemdash}The Hubble Space Telescope Observations, Imaging Data Products, and Mosaics}",
      journal = {\apjs},
         year = 2011,
        month = dec,
       volume = {197},
       number = {2},
          eid = {36},
        pages = {36},
          doi = {10.1088/0067-0049/197/2/36},
archivePrefix = {arXiv},
       eprint = {1105.3754},
 primaryClass = {astro-ph.CO},
       adsurl = {https://ui.adsabs.harvard.edu/abs/2011ApJS..197...36K}
}

@ARTICLE{Carnall2019,
       author = {{Carnall}, A.~C. and {McLure}, R.~J. and {Dunlop}, J.~S. and {Cullen}, F. and {McLeod}, D.~J. and {Wild}, V. and {Johnson}, B.~D. and {Appleby}, S. and {Dav{\'e}}, R. and {Amorin}, R. and {Bolzonella}, M. and {Castellano}, M. and {Cimatti}, A. and {Cucciati}, O. and {Gargiulo}, A. and {Garilli}, B. and {Marchi}, F. and {Pentericci}, L. and {Pozzetti}, L. and {Schreiber}, C. and {Talia}, M. and {Zamorani}, G.},
        title = "{The VANDELS survey: the star-formation histories of massive quiescent galaxies at 1.0 < z < 1.3}",
      journal = {\mnras},
         year = 2019,
        month = nov,
       volume = {490},
       number = {1},
        pages = {417-439},
          doi = {10.1093/mnras/stz2544},
archivePrefix = {arXiv},
       eprint = {1903.11082},
 primaryClass = {astro-ph.GA},
       adsurl = {https://ui.adsabs.harvard.edu/abs/2019MNRAS.490..417C}
}

@ARTICLE{French2023,
       author = {{French}, K. Decker and {Earl}, Nicholas and {Novack}, Annemarie B. and {Pardasani}, Bhavya and {Pillai}, Vismaya R. and {Tripathi}, Akshat and {Verrico}, Margaret E.},
        title = "{Fading AGNs in Poststarburst Galaxies}",
      journal = {\apj},
         year = 2023,
        month = jun,
       volume = {950},
       number = {2},
          eid = {153},
        pages = {153},
          doi = {10.3847/1538-4357/acd249},
archivePrefix = {arXiv},
       eprint = {2304.10419},
 primaryClass = {astro-ph.GA},
       adsurl = {https://ui.adsabs.harvard.edu/abs/2023ApJ...950..153F}
}

@ARTICLE{Hopkins2012AGN,
       author = {{Hopkins}, Philip F.},
        title = "{Dynamical delays between starburst and AGN activity in galaxy nuclei}",
      journal = {\mnras},
         year = 2012,
        month = feb,
       volume = {420},
       number = {1},
        pages = {L8-L12},
          doi = {10.1111/j.1745-3933.2011.01179.x},
archivePrefix = {arXiv},
       eprint = {1101.4230},
 primaryClass = {astro-ph.CO},
       adsurl = {https://ui.adsabs.harvard.edu/abs/2012MNRAS.420L...8H}
}

@ARTICLE{Risaliti1999,
       author = {{Risaliti}, G. and {Maiolino}, R. and {Salvati}, M.},
        title = "{The Distribution of Absorbing Column Densities among Seyfert 2 Galaxies}",
      journal = {\apj},
         year = 1999,
        month = sep,
       volume = {522},
       number = {1},
        pages = {157-164},
          doi = {10.1086/307623},
archivePrefix = {arXiv},
       eprint = {astro-ph/9902377},
 primaryClass = {astro-ph},
       adsurl = {https://ui.adsabs.harvard.edu/abs/1999ApJ...522..157R}
}

@ARTICLE{Lawrence2010,
       author = {{Lawrence}, Andy and {Elvis}, Martin},
        title = "{Misaligned Disks as Obscurers in Active Galaxies}",
      journal = {\apj},
         year = 2010,
        month = may,
       volume = {714},
       number = {1},
        pages = {561-570},
          doi = {10.1088/0004-637X/714/1/561},
       adsurl = {https://ui.adsabs.harvard.edu/abs/2010ApJ...714..561L}
}

@ARTICLE{Hasinger2008,
       author = {{Hasinger}, G.},
        title = "{Absorption properties and evolution of active galactic nuclei}",
      journal = {\aap},
         year = 2008,
        month = nov,
       volume = {490},
       number = {3},
        pages = {905-922},
          doi = {10.1051/0004-6361:200809839},
archivePrefix = {arXiv},
       eprint = {0808.0260},
 primaryClass = {astro-ph},
       adsurl = {https://ui.adsabs.harvard.edu/abs/2008A&A...490..905H}
}

@ARTICLE{Elvis1994,
       author = {{Elvis}, Martin and {Wilkes}, Belinda J. and {McDowell}, Jonathan C. and {Green}, Richard F. and {Bechtold}, Jill and {Willner}, S.~P. and {Oey}, M.~S. and {Polomski}, Elisha and {Cutri}, Roc},
        title = "{Atlas of Quasar Energy Distributions}",
      journal = {\apjs},
         year = 1994,
        month = nov,
       volume = {95},
        pages = {1},
          doi = {10.1086/192093},
       adsurl = {https://ui.adsabs.harvard.edu/abs/1994ApJS...95....1E}
}

@ARTICLE{Kirkpatrick2017,
       author = {{Kirkpatrick}, Allison and {Alberts}, Stacey and {Pope}, Alexandra and {Barro}, Guillermo and {Bonato}, Matteo and {Kocevski}, Dale D. and {P{\'e}rez-Gonz{\'a}lez}, Pablo and {Rieke}, George H. and {Rodr{\'\i}guez-Mu{\~n}oz}, Lucia and {Sajina}, Anna and {Grogin}, Norman A. and {Mantha}, Kameswara Bharadwaj and {Pandya}, Viraj and {Pforr}, Janine and {Salvato}, Mara and {Santini}, Paola},
        title = "{The AGN-Star Formation Connection: Future Prospects with JWST}",
      journal = {\apj},
         year = 2017,
        month = nov,
       volume = {849},
       number = {2},
          eid = {111},
        pages = {111},
          doi = {10.3847/1538-4357/aa911d},
archivePrefix = {arXiv},
       eprint = {1706.09056},
 primaryClass = {astro-ph.GA},
       adsurl = {https://ui.adsabs.harvard.edu/abs/2017ApJ...849..111K}
}

@ARTICLE{Chien2024,
       author = {{Chien}, Tom C.-C. and {Ling}, Chih-Teng and {Goto}, Tomotsugu and {Wu}, Cossas K.-W. and {Kim}, Seong Jin and {Hashimoto}, Tetsuya and {Lin}, Yu-Wei and {Kilerci}, Ece and {Ho}, Simon C.-C. and {Wang}, Po-Ya and {Raquel}, Bjorn Jasper R.},
        title = "{Finding dusty AGNs from the JWST CEERS survey with mid-infrared photometry}",
      journal = {\mnras},
         year = 2024,
        month = jul,
       volume = {532},
       number = {1},
        pages = {719-733},
          doi = {10.1093/mnras/stae1550},
archivePrefix = {arXiv},
       eprint = {2406.14888},
 primaryClass = {astro-ph.GA},
       adsurl = {https://ui.adsabs.harvard.edu/abs/2024MNRAS.532..719C}
}

@ARTICLE{Muzzin2025,
       author = {{Muzzin}, Adam and {Suess}, Katherine A. and {Marchesini}, Danilo and {Robbins}, Luke and {Willott}, Chris J. and {Alberts}, Stacey and {Antwi-Danso}, Jacqueline and {Asada}, Yoshihisa and {Brammer}, Gabriel and {Cutler}, Sam E. and {Iyer}, Kartheik G. and {Labbe}, Ivo and {Martis}, Nicholas S. and {Miller}, Tim B. and {Mitsuhashi}, Ikki and {Pope}, Alexandra and {Sajina}, Anna and {Sarrouh}, Ghassan T.~E. and {Sharma}, Monu and {Stefanon}, Mauro and {Whitaker}, Katherine E. and {Abraham}, Roberto and {Atek}, Hakim and {Bradac}, Marusa and {Berek}, Samantha and {Bezanson}, Rachel and {Brown}, Westley and {Burgasser}, Adam J. and {Chicoine}, Nathalie and {Cloonan}, Aidan P. and {Cooper}, Olivia R. and {Dayal}, Pratika and {de Graaff}, Anna and {Desprez}, Guillaume and {Feldmann}, Robert and {Forrest}, Ben and {Franx}, Marijn and {Fudamoto}, Yoshinobu and {Fujimoto}, Seiji and {Furtak}, Lukas J. and {Glazebrook}, Karl and {Goovaerts}, Ilias and {Greene}, Jenny E. and {Jagga}, Naadiyah and {Jarvis}, William W.~H. and {Kriek}, Mariska and {Khullar}, Gourav and {La Torre}, Valentina and {Leja}, Joel and {Lin}, Jamie and {Lorenz}, Brian and {Lyon}, Daniel and {Markov}, Vladan and {Maseda}, Michael V. and {McConachie}, Ian and {Merchant}, Maya and {Merida}, Rosa M. and {Mowla}, Lamiya and {Myers}, Katherine and {Naidu}, Rohan P. and {Nanayakkara}, Themiya and {Nelson}, Erica J. and {Noirot}, Gael and {Oesch}, Pascal A. and {Omori}, Kiyoaki C. and {Pan}, Richard and {Porraz Barrera}, Natalia and {Price}, Sedona H. and {Ravindranath}, Swara and {Sawicki}, Marcin and {Setton}, David J. and {Smit}, Renske and {Sok}, Visal and {Speagle}, Joshua S. and {Taylor}, Edward N. and {Tan}, Vivian Yun Yan and {Tripodi}, Roberta and {van der Wel}, Arjen and {Perez Vidal}, Edgar and {Wang}, Bingjie and {Weaver}, John R. and {Williams}, Christina C. and {Withers}, Sunna and {Zaidi}, Kumail},
        title = "{MINERVA: A NIRCam Medium Band and MIRI Imaging Survey to Unlock the Hidden Gems of the Distant Universe}",
      journal = {arXiv e-prints},
         year = 2025,
        month = jul,
          eid = {arXiv:2507.19706},
        pages = {arXiv:2507.19706},
          doi = {10.48550/arXiv.2507.19706},
archivePrefix = {arXiv},
       eprint = {2507.19706},
 primaryClass = {astro-ph.GA},
       adsurl = {https://ui.adsabs.harvard.edu/abs/2025arXiv250719706M}
}

@ARTICLE{Belli2019,
       author = {{Belli}, Sirio and {Newman}, Andrew B. and {Ellis}, Richard S.},
        title = "{MOSFIRE Spectroscopy of Quiescent Galaxies at 1.5 < z < 2.5. II. Star Formation Histories and Galaxy Quenching}",
      journal = {\apj},
         year = 2019,
        month = mar,
       volume = {874},
       number = {1},
          eid = {17},
        pages = {17},
          doi = {10.3847/1538-4357/ab07af},
archivePrefix = {arXiv},
       eprint = {1810.00008},
 primaryClass = {astro-ph.GA},
       adsurl = {https://ui.adsabs.harvard.edu/abs/2019ApJ...874...17B}
}



\appendix

\section{Results for the 0.5 < \emph{\MakeLowercase{z}} < 1 bin data}\label{sec:Appendix}

\begin{figure}
\centering
\includegraphics[width=0.99\linewidth]{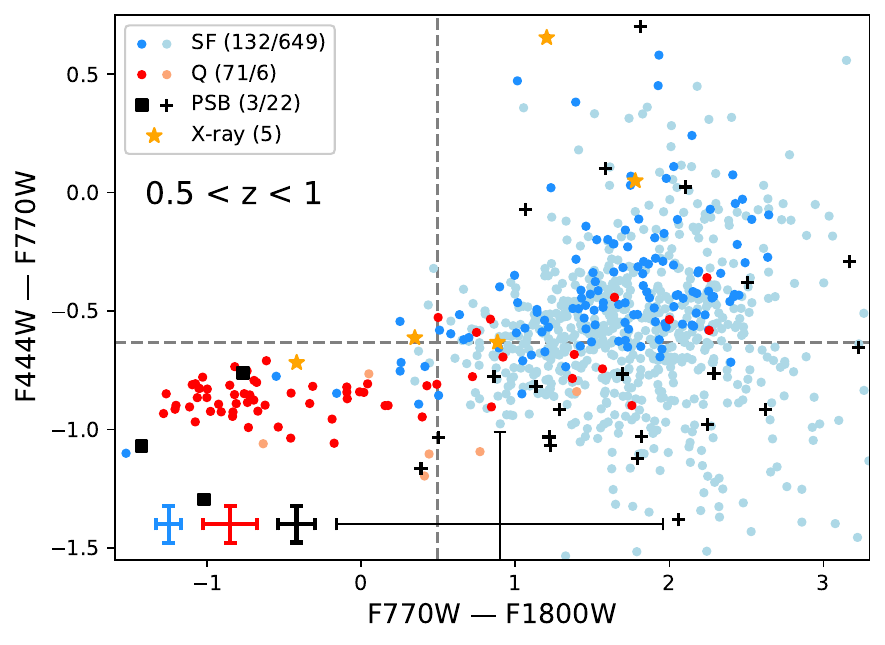}
\caption{Equivalent to Figure \ref{fig:colourcolour} but with the data from the $0.5<z<1$ bin. Comparison of the F770W--F1800W colour and the F444W--F770W colour for the star-forming, quiescent, and PSB populations. The populations are separated by a mass split at $10^{10}\textrm{\;M}_\odot$, with the black squares being the massive PSBs and the black crosses being the low-mass PSBs. The vertical and horizontal grey dashed lines indicate the respective separation location of the populations in F770W--F1800W and F444W--F770W (both based on the KDE minima of each of the massive star-forming and quiescent colour distributions), with galaxies exhibiting a colour redder (right) of the vertical cut considered to have a \emph{MIR excess}, and those with a colour bluer (left) of the vertical cut considered to have a \emph{MIR non-excess}. Galaxies that have X-ray emission detected by \emph{Chandra} imaging are denoted by yellow stars. At the bottom, median 1$\sigma$ colour errors of the populations are shown, corresponding to the massive star-forming, massive quiescent, massive PSB, and low-mass PSB, from left to right.}
\label{fig:colourcolour_appndx}
\end{figure}

\begin{figure*}
\centering
\includegraphics[width=0.9\linewidth]{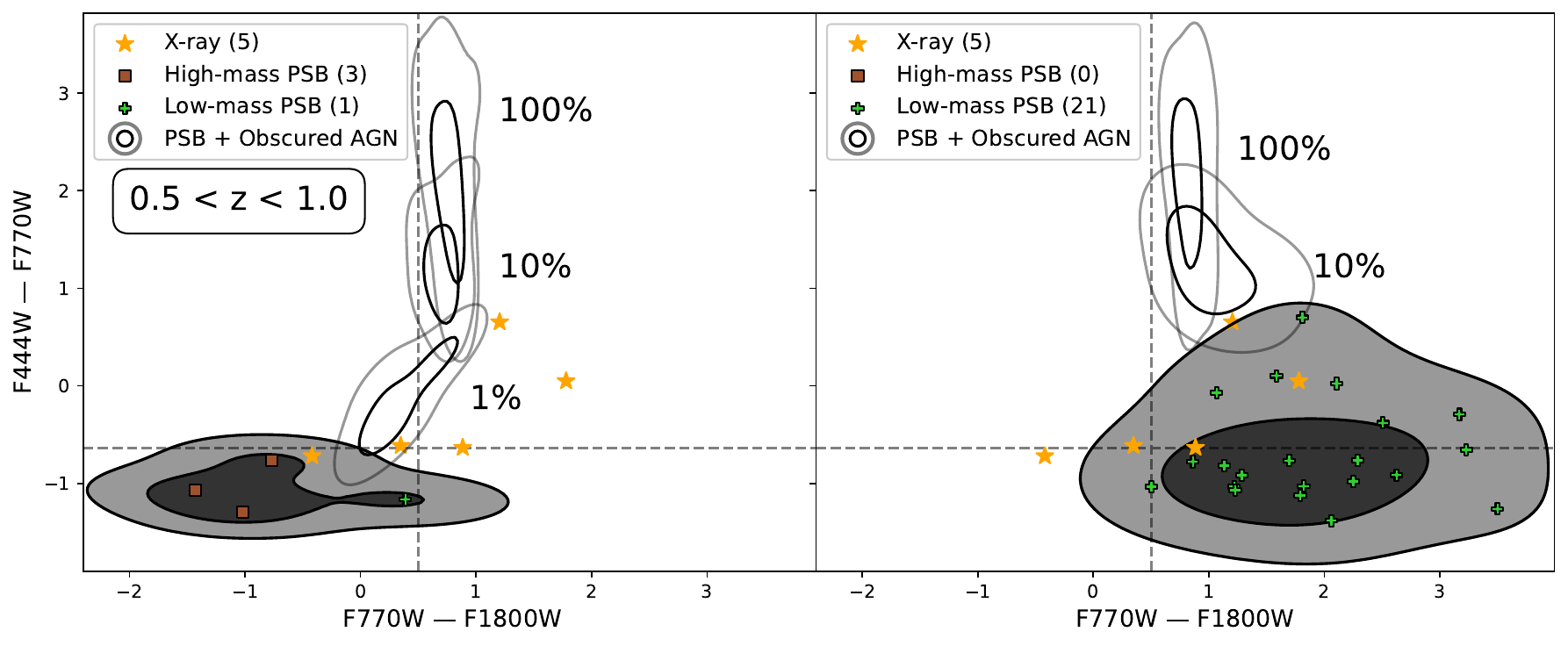}
\caption{Equivalent to Figure \ref{fig:agn_probs} but with the data from the $0.5<z<1$ bin. Colour-colour locations and probability distributions of PSBs, and how their colours change with the addition of an AGN template accreting at various Eddington ratios. The left panel shows the fiducial location of all MIR non-excess PSBs (F770W--F1800W values less than the vertical dashed line), and the right panel shows the fiducial location of all MIR excess PSBs (with high-mass PSBs denoted as brown squares and low-mass as green crosses). In each panel the black and grey contours indicate the 1$\sigma$ and 2$\sigma$ probability distributions, respectively, with the filled black and grey contours indicating the same but for the fiducial PSB population. The clear contours in each panel indicate the colour location of PSBs when AGN with viewing angles of $50^\circ \, \text{to} \, 90^\circ$ are added, indicating the effect of an obscured AGN. For the MIR non-excess population, contours indicating the addition of AGN accreting at $1\%$, $10\%$, and $100\%$ Eddington are shown, while for the MIR excess population only $10\%$ and $100\%$ are shown due to the similarity of the $1\%$ contours with the fiducial population. Additionally, the locations of all sources with secure X-ray detection are denoted by yellow stars, regardless of their SC classification.}
\label{fig:agn_tracks_appndx}
\end{figure*}

As mentioned in Section \ref{sec:Colour_sep}, we omit the results from the $0.5<z<1$ bin for the sake of brevity and visual clarity due to the limited numbers of high-mass PSBs. However, we noted that this excluded bin is consistent with trends from the other bins and does not alter our conclusions. The equivalent plots to Figures \ref{fig:colourcolour} and \ref{fig:agn_probs} for the $0.5<z<1$ bin are shown in Figures \ref{fig:colourcolour_appndx} and \ref{fig:agn_tracks_appndx}, respectively.

Figure \ref{fig:colourcolour_appndx} shows the separation between the massive star-forming and quiescent galaxies in the F770W--F1800W/F444W--F770W colour-colour space, with quiescents residing in the lower-left and star-forming in the upper right. The high-mass PSBs are exclusively in the lower left, while the low-mass PSBs all have elevated F770W--F1800W colours, and a range of F444W--F770W colours. The few massive PSBs in this bin have similar colours to the quiescent population but also a larger scatter, especially when comparing to the populations in Figure \ref{fig:colourcolour}. Visual inspection of the massive PSB with the lowest F444W--F770W colour shows that it has a close companion that is likely contaminating its flux values, and thus could be the explanation for its offset.

Figure \ref{fig:agn_tracks_appndx} shows the effect of adding obscured AGN templates to the fiducial SEDs of the PSB population, with the MIR non-excess and MIR excess on the left and right panel, respectively. The same results that were shown in Figure \ref{fig:agn_probs} are also found here, with the MIR non-excess population (effectively the high-mass PSBs) being significantly (2$\sigma$) different in colour space with AGN models of $\sim1\%$ Eddington ratios, while the MIR excess population (effectively the low-mass PSBs) being significantly different with models of $\sim10\%$ Eddington ratios. The probability distributions are more constrained in F770W--F1800W than those in Figure \ref{fig:agn_probs}, due to those higher redshift bins probing restframe wavelengths that are still slightly affected by dust attenuation. At $0.5<z<1$, the colours are almost entirely unaffected, leading to near identical tracks between all viewing angles.

In the right panel of Figure \ref{fig:agn_tracks_appndx}, a higher portion of low-mass PSBs seem to reside in the same colour space as the X-ray sources. However, as is indicated by the increased median $1\sigma$ errors in Figure \ref{fig:colourcolour_appndx}, the mass completeness in this redshift range allows for objects of lower mass, and in turn increased flux errors. Indeed, the low-mass PSB population with elevated F444W--F770W values is dominated by the lowest-mass galaxies.

\bsp	
\label{lastpage}
\end{document}